\documentstyle[
   aps,
   pre,
   preprint,
   tighten,
   epsf,
]{revtex}

\epsfclipon

\begin{document}

\draft

\title{
            UV continuum emission and diagnostics\protect\\
         of hydrogen-containing non-equilibrium plasmas
}
\author{Boris P. Lavrov and Alexei S. Melnikov}
\address{Faculty of Physics, St.-Petersburg State University, 198904, Russian 
Federation}
\author{Marko K\"aning and J\"urgen R\"opcke}
\address{Institute for Low Temperature Plasma Physics, 17489 Greifswald,
Federal Republic of Germany}
\date{\today}
\maketitle

\begin{abstract}
For the first time the emission of the radiative dissociation continuum of the 
hydrogen molecule ($a^{3}\Sigma_{g}^{+} \to b^{3}\Sigma_{u}^{+}$ electronic 
transition) is proposed to be used as a source of information for the 
spectroscopic diagnostics of non-equilibrium plasmas. The detailed analysis of 
excitation-deactivation kinetics, rate constants of various collisional and 
radiative transitions and fitting procedures made it possible to develop two 
new methods of diagnostics of: (1) the ground $X^{1}\Sigma_{g}^{+}$ state 
vibrational temperature $T_{\text{vib}}$ from the relative intensity 
distribution, and (2) the rate of electron impact dissociation 
$(d[\mbox{H$_{2}$}]/dt)_{\text{diss}}$ from the absolute intensity of the 
continuum. The known method of determination of $T_{\text{vib}}$ from relative 
intensities of Fulcher-$\alpha$ bands was seriously corrected and simplified 
due to the revision of $d \to a$ transition probabilities and cross sections 
of $d \gets X$ electron impact excitation. General considerations are 
illustrated with examples of experiments in pure hydrogen capillary-arc and 
H$_{2}$+Ar microwave discharges. 
In pure H$_{2}$ plasma the values of $T_{\text{vib}}$ obtained by two 
independent methods are in rather good accordance ($T_{\text{vib}}$=3000-5000 
K).
In the H$_{2}$+Ar microwave plasma it was observed for the first time that the 
shape of the continuum depends on the ratio of the mixture components. 
Absorption measurements of the population of the 3s$^2$3p$^5$4s levels of Ar 
together with certain computer simulations showed that the Ar*$\to$H$_{2}$ 
excitation transfer plays a significant role. In our typical conditions (power 
flux: 4~W cm$^{-2}$, pressure $p=0.5$~mbar, H$_{2}$:Ar=1:1) the following 
values were obtained for the microwave discharge: 
$(d[\mbox{H$_{2}$}]/dt)_{\text{diss}}\approx 2.5 - 5 \cdot 10^{17}$ cm$^{-3}$ 
s$^{-1}$. The contribution of the excitation transfer is about 10-30\% of the 
total population of the $a^{3}\Sigma_{g}^{+}$ state.

\end{abstract}


\narrowtext
% body of paper here
\narrowtext

\section{Introduction}

Experimental and theoretical studies of the emission of the H$_{2}$ radiative 
dissociation continuum ($a^{3}\Sigma_{g}^{+} \to b^{3}\Sigma_{u}^{+}$ 
electronic transition) have a long and interesting history. They were 
stimulated by molecular spectroscopy and quantum chemistry 
\cite{JC39,FH65,TF72,LFSM85,KGDP86,LLP88,LLP89}, investigations of hydrogen 
containing plasmas \cite{Zaidel,LP85,LP88} and astrophysics 
\cite{Doyle68,SC72}. Since the first work of Houtermans \cite{Hm60} the 
$a^{3}\Sigma_{g}^{+} \to b^{3}\Sigma_{u}^{+}$ transition has been studied as a 
possible laser system (see \cite{CM73,GY78,GP80,BGP81} and Refs.~therein). It 
is widely used in UV and VUV light sources 
\cite{Zaidel,GS79,h2lamp1,h2lamp2,LT86}. 

In spite of numerous studies, the H$_{2}$ continuum emission was never used or 
even proposed to be used for a determination of plasma parameters, i.e., for 
spectroscopic plasma diagnostics. It seems to be rather strange because, on 
one hand, diagnostics of hydrogen-containing plasmas is not trivial and every 
opportunity should be tried; on the other hand, emission spectroscopy due to 
its passive nature and instrumental simplicity is of primary interest for 
various studies in plasma physics and industrial applications. Moreover the 
visible spectrum of hydrogen plasma gives us in principle only three 
qualitatively different sources of information: (1) atomic lines of Balmer 
series, (2) band (actually multi-line) spectrum of H$_{2}$, and (3) the 
continuum. First two are successfully used in plasma diagnostics, while the 
third one was never even tried. Previous investigators only measured the 
continuum intensity and sometimes made certain calculations for comparison 
with an experiment \cite{LP85,LP88}. It should be underlined that those 
calculations are solutions of so-called direct problem of spectroscopy (from 
known plasma parameters to observable spectrum). 

It is obvious that any observable property of plasma is somehow connected with 
values of plasma parameters and therefore in principle may be tried to be used 
as a diagnostic tool. However actually, as a rule, one observable quantity is 
determined by several different plasma parameters. Then the key point is to 
study the informational content of experimental data and to find the proper 
way of data analysis. The last one is always an inversion problem (from an 
observed spectrum to unknown plasma parameters) having its own specifics (see 
below section \ref{sec:Theory}).

The main goal of the present work was to investigate the possibility to use 
H$_{2}$ continuum intensity for diagnostics of non-equilibrium plasmas. Our 
decision to make the first steps in this direction was based on following 
\textit{a priory} considerations. The experimental data on the continuum 
intensity may be considered as two different sources of information: the 
relative spectral distribution and the total absolute intensity of the 
continuum. 

The first one is directly connected with the population density distribution 
over vibrational levels of the $a^{3}\Sigma_{g}^{+}$ state. In low pressure 
plasmas the levels are populated predominantly by direct electron impact 
excitation and the deactivation is mainly due to spontaneous emission 
\cite{LP85,LP88}. In this case the relation between the vibrational population 
density distributions in excited and ground electronic states should be 
obtained within corona-like models. So the shape of the continuum should be 
usable for a determination of the vibrational density distribution (or at 
least the vibrational temperature) of the ground state of H$_{2}$.

On the other hand every act of the continuum emission leads to a dissociation 
of a hydrogen molecule. Therefore, an absolute value of the continuum 
intensity is directly connected with the rate of radiative dissociation via 
the $a^{3}\Sigma_{g}^{+}$ state and also may be used for an estimation of the 
total rate of electron impact dissociation. 

Our intention was to investigate both opportunities and to check them 
experimentally.\footnote{Some very preliminary results were already reported 
in \cite{FrontII1}.} Therefore the present work includes two closely connected 
tasks (development of new methods and their test in certain real plasma 
conditions) and two groups of results connected with the proposed methods and 
with non-equilibrium plasma of the discharges under the study. 

The first part is based on the analysis and description of: 
\begin{enumerate}
\item excitation-deactivation kinetics and derivation of suitable formulas 
(Sections \ref{sec:ContIntPop}, \ref{sec:ExcDeacBalEq}, \ref{sec:DetTv}, 
\ref{sec:dissrate});
\item rate constants of various collisional and radiative transitions to 
establish certain sets of recommended data (Section \ref{sec:Constants});
\item experimental technique reproducible by other researchers (Section 
\ref{sec:experiment});
\item various ways of data processing (Section \ref{sec:ContIntPop} and 
throughout the paper).
\end{enumerate}

That made it possible to develop two new methods of diagnostics of: (1) the 
ground $X^{1}\Sigma_{g}^{+}$ state vibrational temperature $T_{\text{vib}}$ 
from the relative intensity distribution, and (2) the rate of electron impact 
dissociation from the absolute intensity of the continuum. The known method of 
determination of $T_{\text{vib}}$ from relative intensities of 
Fulcher-$\alpha$ bands was seriously corrected and simplified due to the 
revision of $d \to a$ transition probabilities and cross sections of $d \gets 
X$ electron impact excitation.

For tests of the methods we performed the studies of emission spectra of two 
quite different plasma sources: (1) pure hydrogen dc capillary-arc (Section 
\ref{sec:dc}), and (2) H$_{2}$+Ar microwave discharges (Section \ref{sec:mw}).

The first one is known to produce bright pure hydrogen spectrum and was 
already investigated spectroscopically \cite{LP85,LP88,LT86,LT82}\footnote{Our 
previous attempt to obtain $T_{\text{vib}}$ from Fulcher-$\alpha$ band 
intensities in \cite{LP85} almost failed, because of insufficient precision of 
intensity measurements (errors $\approx$ 10\%).}. Therefore it was chosen as 
an example of a simple model system for determination of $T_{\text{vib}}$ by 
two different methods presented above.

The second one has been specially designed for basic research on molecular 
microwave discharges mainly based on spectroscopic diagnostic methods 
\cite{Ohl92b,RKL98}. Microwave plasma reactors of this planar type have been 
used for various plasma technological applications 
\cite{RKL98,ORS93,ROS93,OSRKPS95}. Frequently H$_{2}$ and Ar are used as main 
components of the feed gas mixture leading to an essential interest on H$_{2}$ 
dissociation processes and on influences of Ar admixture on H$_{2}$ emission. 

The results are summarized and discussed in the conclusion.

\section{Intensities of H$_{2}$ bands and vibrational temperature}
\label{sec:Theory}

\subsection{Continuum intensity and \protect\\population of upper levels}
\label{sec:ContIntPop}

The dissociation continuum of the hydrogen molecule is caused by spontaneous 
transitions from the upper bound $a^{3}\Sigma_{g}^{+}$ electronic state to the 
lower $b^{3}\Sigma_{u}^{+}$ repulsive state. The spectral distribution 
$I_{ab}(\lambda)$ of the continuum intensity (number of quanta emitted within 
unit range of wavelengths per unit volume and per second in all directions) is 
related to the population densities $N_{av'}$ of the excited 
$a^{3}\Sigma_{g}^{+},v'$ vibronic levels, as
\begin{equation}
I_{ab}(\lambda) = \sum\limits_{v'}
N_{av'} \cdot A_b^{av'}(\lambda),
\label{eq:Iab}
\end{equation}
where $\lambda$ -- wavelength, $v'$ -- vibrational quantum number,
\begin{equation}
 N_{av'} = \sum\limits_{N'} N_{av'N'},
\label{eq:Nav}
\end{equation}
the total population density of the upper vibronic state,  $N_{av'N'}$ -- 
populations of $a^{3}\Sigma_{g}^{+},v',N'$ electronic-vibro-rotational levels 
(their very small triplet splitting may be neglected), $N'$ -- rotational 
quantum number of the total angular momentum excluding electron spin, and 
$A_b^{av'}(\lambda)$ -- spectral distribution of the spontaneous emission 
transition probability (Einstein coefficients for various wavelengths of 
$a^{3}\Sigma_{g}^{+},v' \to b^{3}\Sigma_{u}^{+}$ transitions). We have 
neglected the rather small effect of vibro-rotation interaction and assume the 
transition probabilities being independent on $N'$.

\begin{figure}
 \epsfxsize=8.6cm \epsfbox{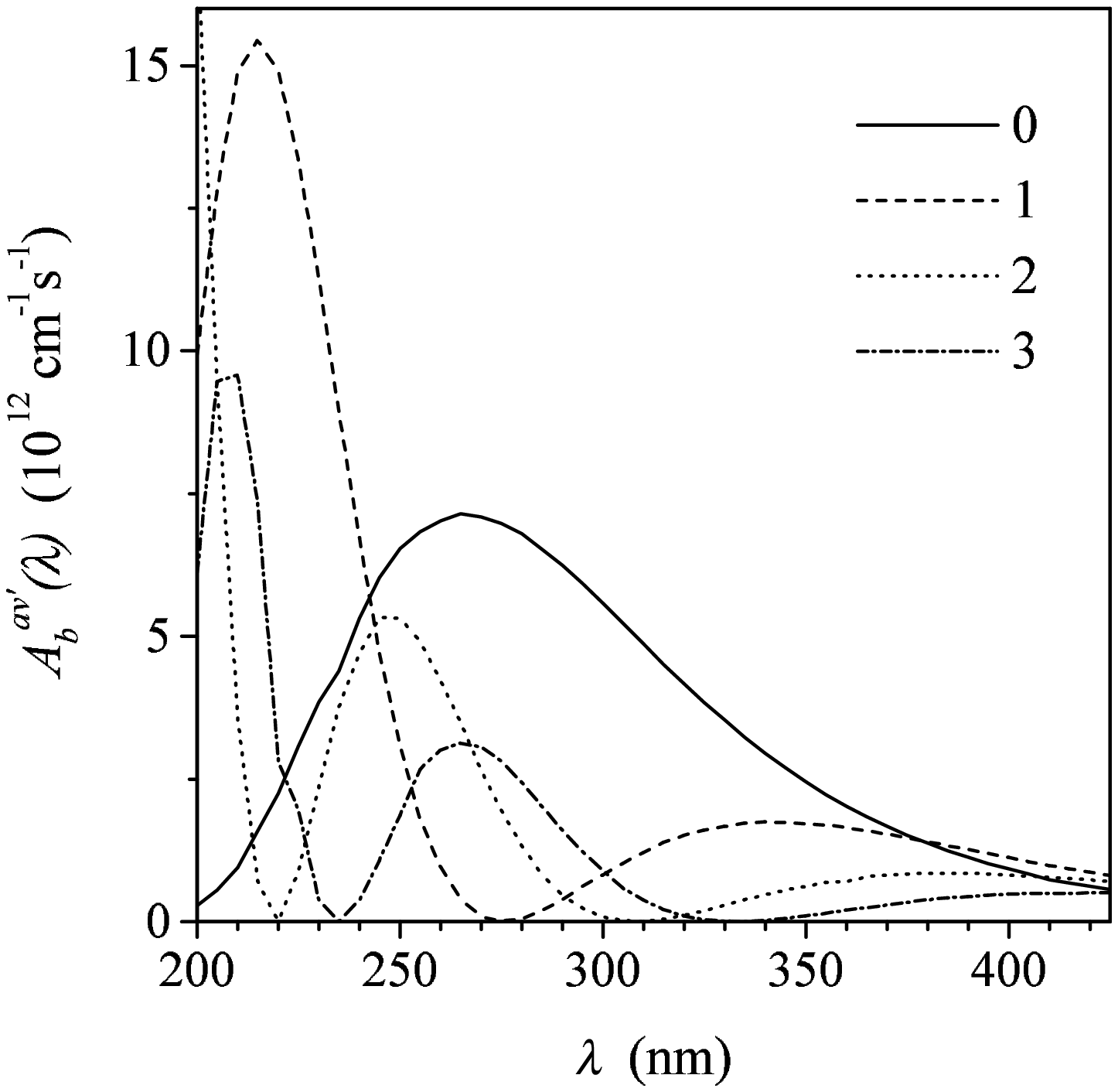}
 \caption{Spectral distribution of $a^{3}\Sigma_{g}^{+},v' \to 
b^{3}\Sigma_{u}^{+}$ spontaneous emission transition probabilities for various 
initial vibronic levels $v'=0-3$ (curves $0-3$) of the H$_{2}$ molecule 
\protect\cite{LLP88,LLP89}.}
 \label{fig:1}
\end{figure}

For H$_{2}$ molecule the values of $A_b^{av'}(\lambda)$ should be considered 
as well-known nowadays because they have been calculated in several works 
\cite{JC39,KGDP86,LLP88,LLP89,CM73} and checked experimentally 
\cite{JC39,LFSM85}, being in rather good agreement \cite{LLP89}. We used the 
results of \cite{LLP88} shown in Table~\ref{tab:Alv}.\footnote{Those are the 
only data presented in table form suitable for applications. The data were 
never published in open press, but deposited in VINITI (Institute of 
Scientific and Technical Information of USSR).} In Fig.~\ref{fig:1} the data 
are visualized for the transitions from $a^{3}\Sigma_{g}^{+},v'$=0-3 levels 
for $\lambda=200-450$~nm. One may see that the probability of transitions from 
various vibronic levels has maxima in different wavelength regions due to 
oscillatory behavior of the vibrational wave functions of the upper state. So 
the spectral distribution of the continuum intensity $I_{ab}(\lambda)$ may 
have a certain informational content about the population density distribution 
over vibrational levels of the $a^{3}\Sigma_{g}^{+}$ electronic state. In 
principle, measurements of the intensity $I_{ab}(\lambda)$ may be used for the 
determination of the populations $N_{av'}$ in relative or even in absolute 
scale by Eq.~(\ref{eq:Iab}). 

It should be noted that even this very first stage of the data processing 
belongs to the class of so-called reverse (or inversion) problems, which are 
known to be not well-posed (sometimes they are called ``ill-posed''). The 
final result -- a certain set of the populations $N_{av'}$ -- depends (in 
principle) not only on the values of the actual populations in plasma, but may 
be influenced by so-called \textit{a priory} information: the number of 
adjusted parameters ($N_{av'}$), the number of experimental data points, the 
experimental errors and even the algorithm of data processing.

In the present work we used for the solution of this and other reverse 
problems the least-square fitting, i.e., the minimization of the functional
\begin{equation}
\chi^2 = \frac{1}{l-m}\
\sum\limits_{i=1}^{l}
\left(
\frac{Y_i^{\text{calc}}(a_1,\dots,a_m)
- Y_i^{\text{expt}}}{\sigma_i}
\right)^2
\label{eq:chi2}\\
\end{equation}
in the multidimensional space of parameters $a_1,\dots,a_m$ used in a 
theoretical model. Here $Y_i^{\text{calc}}(a_1,\dots,a_m)$, 
$Y_i^{\text{expt.}}$ -- calculated and experimental data, $l$ -- total number 
of data points, $m$ -- number of adjusted parameters, $\sigma_i$ -- standard 
deviations of the experimental data. The standard deviations of optimal values 
of the adjusted parameters may be estimated as square roots of diagonal terms 
of the covariance matrix \cite{Hudson64}.

We used a minimization algorithm analogous to a linear regression 
\cite{Hudson64} with the following substitutions: $Y_i^{\text{expt}}\equiv 
I_{ab}^{\text{expt}}(\lambda_i)$ from experiment, $Y_i^{\text{calc}}\equiv 
I_{ab}(\lambda_i)$ from Eq.~(\ref{eq:Iab}), $\sigma_i\equiv 0.07\cdot 
I_{ab}^{\text{expt}}(\lambda_i)$ corresponding to 7\% random experimental 
error of the measurement, $m\equiv v'_{\text{max}} +1$, $\lambda_i$ was varied 
in the range $225-400$~nm. Experimental errors were estimated by averaging of 
several measurements for the same conditions in plasma. They are mainly caused 
by random noise, non-sufficient stability of the discharges during the 
experiments and restricted reproducibility. The value of $v'_{\text{max}}$ had 
to be adjusted in such a way that reliable results can be obtained from the 
minimization procedure. The values of $A_b^{av'}(\lambda)$ from 
Table~\ref{tab:Alv} were interpolated by cubic splines.

\subsection{Excitation-deactivation balance equation
\label{sec:ExcDeacBalEq}
}

The populations of $N_{av'}$ may be sometimes interesting themselves (see the 
examples below). But on the other hand, in non-equilibrium plasmas the 
rovibronic level populations in excited electronic states are connected with 
those in the ground electronic state by the excitation-deactivation balance 
equation for the electronically excited $n',v',N'$ rovibronic levels \cite{L84}
\widetext
\begin{equation}
\sum\limits_{vN} N_{XvN} \cdot n_e
\cdot \alpha_{XvN}^{n'v'N'} +
I_{n'v'N'}(\alpha,\beta,\gamma,\dots) =
N_{n'v'N'}
\left(
\frac{1}{\tau_{n'v'N'}^{\text{rad}}}
+
\frac{1}{\tau_{n'v'N'}^{\text{coll}}}
\right),
\label{eq:Balance}
\end{equation}
\narrowtext
where $I_{n'v'N'}(\alpha,\beta,\gamma,\dots)$ -- the rate of secondary 
excitation processes (cascades, recombination and so forth); 
$\tau_{n'v'N'}^{\text{rad}}$ -- radiative lifetime of the $n'v'N'$ level; 
$\tau_{n'v'N'}^{\text{coll}}$ -- effective lifetime describing the decay of 
the $n'v'N'$ level due to quenching in collisions of the excited molecule with 
electrons, atoms and unexcited molecules. The rate coefficient of direct 
electron impact excitation
\widetext
\begin{equation}
\alpha_{XvN}^{n'v'N'} = \langle \sigma{\mathbf v\rangle}_{XvN}^{n'v'N'} =
\int\nolimits_{\varepsilon_{th}}^{\infty}
\sqrt{\frac{2\varepsilon}{m}}\
\sigma_{XvN}^{n'v'N'}(\varepsilon)\
F(\varepsilon)\ d\varepsilon,
\label{eq:alpha}
\end{equation}
\narrowtext
where $\sigma_{XvN}^{n'v'N'}(\varepsilon)$ -- corresponding cross section; 
$\mathbf{v}, \varepsilon$ -- velocity and corresponding energy of an incident 
electron; $\varepsilon_{th}$ -- threshold energy of the $n',v',N' \leftarrow 
X,v,N$ transition; $F(\varepsilon)$ -- the electron energy distribution 
function.

For plasma diagnostics this most simple limit case (often called corona or 
corona-like model) is that of low-pressure plasmas with small input power. 
Under such conditions the levels are predominantly excited by direct electron 
impact excitation and their decay is mainly due to spontaneous 
emission.\footnote{An additional population of the $a^{3}\Sigma_{g}^{+}$ state 
due to cascades from $e^{3}\Sigma_{g}^{+}$ and $d^{3}\Pi_{u}^{-}$ states was 
taken into account in \cite{LP88} and found to be less than 20\%. The 
radiative and collisional quenching lifetimes are estimated to be equal at 3 
mbar for $a^{3}\Sigma_{g}^{+}$ and at 0.75 mbar for $d^{3}\Pi_{u}^{-}$ states 
(see \cite{LP88,BLMPYY90}).}
Than the second terms in right and left hand parts of Eq.~(\ref{eq:Balance}) 
may be omitted.

It may be seen from Eqs.~(\ref{eq:Nav}) and (\ref{eq:chi2}) that even in the 
most favorable case for plasma diagnostics the populations $N_{n'v'N'}$ depend 
not only on the distribution $N_{XvN}$. Radiative transition probabilities, 
lifetimes and electron impact cross sections for single rovibronic levels and 
transitions should be known as well as the electron energy distribution 
function. Moreover the determination of $N_{XvN}$ from measured $N_{n'v'N'}$ 
and with the certain system of equations (\ref{eq:Nav}) for various $v'$ and 
$N'$ lead to a reverse problem.

\subsection{Constants of elementary processes}
\label{sec:Constants}

Further simplification of the model may be achieved by detailed analysis of 
radiative and collision transition probabilities and lifetimes.
We are interested in determination of the vibrational temperature 
$T_{\text{vib}}$ from the continuum intensity ($a^{3}\Sigma_{g}^{+} \to 
b^{3}\Sigma_{u}^{+}$) and from the intensities of Q-branch lines of 
Fulcher-$\alpha$ bands ($d^{3}\Pi_{u}^{-} \to 
a^{3}\Sigma_{g}^{+}$).\footnote{The method of $T_{\text{vib}}$ derivation from 
Fulcher-$\alpha$ bands was originally proposed in \cite{LP85}. It included 
measurements of Q-line intensities, $F(\varepsilon)$ and gas temperature $T$. 
It was based on rather old data about radiative lifetimes (see 
\cite{BLMPYY90}) and the cross sections of the $d \gets X$ electron impact 
excitation from \cite{LOU81}. Our analysis show that the data should be 
revised and the method sufficiently simplified (see Eqs.~(\ref{eq:SlvTv}), 
(\ref{eq:Ndv}) and the discussion).}
Therefore we have to analyze all available data about electron impact 
excitation and spontaneous decay of the upper levels.

The transition probabilities for $d^{3}\Pi_{u}^{-},v',N'$=1 $\to$ 
$a^{3}\Sigma_{g}^{+},v'',N''$=1 transitions [Q1 lines of $(v'-v'')$ 
Fulcher-$\alpha$ bands] were obtained semi-empirically in the framework of the 
adiabatic approximation with corresponding dipole moment obtained in 
\cite{LP89} from experimental data about wave numbers, branching ratios in 
$v''$-progressions and radiative lifetimes. The results for the first seven 
diagonal ($v'$=$v''$=$v$) bands are presented in Table~\ref{tab:FtA} together 
with semi-empirically predicted \cite{LP89} and experimental values of 
radiative lifetimes \cite{BLMPYY90}.

One may see a noticeable discrepancy between semi-empirical and experimental 
lifetimes of $d^{3}\Pi_{u}^{-},v'$=4-6, $N'$=1 rovibronic levels due to 
non-adiabatic effects neglected in the adiabatic approximation (see also 
\cite{LP88,BLMPYY90}). Therefore, only the first four diagonal bands may be 
used for the determination of $T_{\text{vib}}$. On the other hand this effect 
did not influence the values of the rotational temperatures derived from the 
populations of $d^{3}\Pi_{u}^{-},v' >$3 levels in \cite{AKKKLOR96}. It may be 
considered as an indication, that the rate of the additional decay, associated 
with a non-adiabatic coupling of the vibronic levels lying above the H($n$=1) 
+ H($n$=2) dissociation limit, is almost independent on $N'$ (here $n$ is the 
principle quantum number). It seems quite reasonable, if the perturbation is 
due to an electron-rotational interaction with a continuum  (the hypothesis 
proposed in \cite{L88}). Because of the lack of information about 
$N'$--dependences of the transition probabilities and the lifetimes they will 
be considered herein as independent on the rotational quantum number in 
accordance with an adiabatic approximation.

The radiative lifetimes of $a^{3}\Sigma_{g}^{+},v'$ vibronic states of H$_{2}$ 
have been studied in numerous works both experimentally 
\cite{FH65,TF72,SC72,IR71,KRI75,MK79} and by \textit{ab initio} calculations 
\cite{JC39,KGDP86}. All available data are collected in Table~\ref{tab:tv}. 
One may see, that: 
\begin{list}{}{}
\item[(1)] Nothing is known about $N'$--dependences of the lifetimes, so we 
again have to assume independence on $N'$. 
\item[(2)] Experimental data are obtained only for low vibronic states. 
\item[(3)] The experimental and calculated data are in very good accordance 
for $v'=0-2$, but for higher $v'$ only \textit{ab initio} data are available.
\end{list}
To be in consistence with the transition probabilities listed in 
Table~\ref{tab:Alv} we have used values of $\tau_{av'}$ marked in 
Table~\ref{tab:tv} as ``p.w.'' -- present work. The lifetimes for $v'=0-2$ 
were obtained by the integration of the tabulated values of the transition 
probabilities from Table~\ref{tab:Alv}. The values labeled with an asterisk 
for $v'=3-6$ were obtained by extrapolating the lifetimes with the help of a 
third order polynomial fit of the data from \cite{KGDP86} and a constant 
factor obtained from the average value of the ratios with our values for 
$v'=0-2$.\footnote{This was necessary because for $v' >$2 one can see in 
Table~\ref{tab:Alv} that some quantity of transition probability resides below 
the lower table margin of $\lambda=160$~nm. Therefore, the calculation of the 
$\tau_{av'}$ from the tabulated $A_b^{av'}(\lambda)$ values would lead to an 
overestimation of the lifetimes.}

The dependences of $d^{3}\Pi_{u}^{-},v',N' \leftarrow 
X^{1}\Sigma_{g}^{+}$,0,$N$ electron impact excitation on the incident electron 
energy have been studied in \cite{LOU81,MDH76,BN76}. They have a normal form 
for singlet-triplet transitions: a sharp increase from the threshold to the 
maximum and a rather sharp decrease for higher energies. The data of 
\cite{LOU81,MDH76} are in good accordance and in \cite{DLP86} they were used 
for the extraction of the cross sections for different $v'$ from rate 
coefficients and $F(\varepsilon)$ measured in \cite{LP85}. The relative cross 
sections $\sigma^{dv'1}_{X01}/\sigma^{d21}_{X01} $ in maximum are shown in 
Table~\ref{tab:2} together with corresponding ratios of the Franck-Condon 
factors calculated in \cite{LMPT90}. One may see noticeable deviation of 
measured values from those calculated in the Franck-Condon approximation. In 
\cite{LOU81} this discrepancy was interpreted as an evidence of a remarkable 
dependence of the scattering amplitude on the internuclear distance 
(non-Franck-Condon effect). Later the data from \cite{LP85} have been used for 
the determination of the complete set of the cross sections for various $v,v'$ 
and $N,N'$ \cite{DLP86}.

It is important to take into account that the relative cross sections in 
\cite{LP85,LOU81,DLP86} are based on radiative transition probabilities and 
lifetimes obtained in adiabatic approximation (almost the same as in 
Table~\ref{tab:2}). As we already mentioned above the lifetimes of $v'=4-6$ 
are in contradiction with such extrapolation. New data about $\tau_{dv'1}$ and 
$A^{dv'1}_{av''1}$ from Table~\ref{tab:FtA} should be used for the 
determination of the cross sections from the line intensities measured in 
\cite{LP85}. We made the recalculation and got the data presented in the sixth 
column of Table~\ref{tab:2}. One may see that they are in rather good 
agreement with corresponding ratios of Franck-Condon factors. It means that 
the dependence of the scattering amplitude on the internuclear distance is 
negligible in the case of the $d^{3}\Pi_{u}^{-} \gets X^{1}\Sigma_{g}^{+}$ 
electron impact excitation of H$_{2}$. This is in accordance with recent 
\textit{ab initio} calculations \cite{LMBM91}.

Nothing is known, so far as we know, about the cross sections for the 
$a^{3}\Sigma_{g}^{+},v',N' \gets X^{1}\Sigma_{g}^{+},v,N$ excitation. 
Therefore, for both $d^{3}\Pi_{u}^{-} \gets X^{1}\Sigma_{g}^{+}$ and 
$a^{3}\Sigma_{g}^{+} \gets X^{1}\Sigma_{g}^{+}$ transitions we assume that the 
rate coefficients of the electron impact excitation have non-zero values only 
for $N'=N$ rovibronic transitions. They are considered as independent on $N$ 
and proportional to corresponding Franck-Condon factors, i.e.,
\widetext
\begin{eqnarray}
\alpha_{XvN}^{n'v'N'}&\approx&\alpha_{Xv}^{n'v'} \propto Q_{Xv}^{n'v'}, \qquad
\mbox{when}\qquad \Delta N'\equiv
N'-N=0\\
\mbox{and}\quad
\alpha_{XvN}^{n'v'N'}&=&0\hspace{28mm}
\mbox{for}\hspace{10.5mm} \Delta N'\neq 0
\label{eq:alpha2}
\end{eqnarray}
\narrowtext
thus neglecting the momentum transfer in electron impact excitation 
\cite{LOU81,DL88} and the rather small dependence of the rate coefficients on 
$F(\varepsilon)$. We used the only values of Franck-Condon factors for 
$d^{3}\Pi_{u}^{-} \gets X^{1}\Sigma_{g}^{+}$ and $a^{3}\Sigma_{g}^{+} \gets 
X^{1}\Sigma_{g}^{+}$ transitions which had been calculated in \cite{LMPT90} 
(Tables~\ref{tab:FCF1} and~\ref{tab:FCF2}).\footnote{Although those data 
should in principle be available by order from Russia, we decided to include 
them into the article for completeness of the data set on elementary processes 
involved and for convenience of potential users of our methods.
}

\subsection{Determination of vibrational temperature}
\label{sec:DetTv}

Vibrational and rotational energies are not well separated in the ground 
electronic state of H$_{2}$ because of the small mass of the nuclei. Just to 
have a certain ability to characterize the distribution over vibrational 
levels we introduce the vibrational temperature by
\widetext
\begin{equation}
N_{Xv} =
\sum\limits_N N_{XvN} =\sum\limits_N
\frac{\mbox{[H$_{2}$]}
\cdot
\exp{\Bigl(-\frac{\Delta
E_{Xv}}{kT_{\text{vib}}}\Bigr)}
\cdot
(2N+1)
\cdot
\exp{\Bigl(-\frac{\Delta E_{XvN}}{kT}\Bigr)}
}
{
\sum\limits_{v}
\exp{\Bigl(-\frac{\Delta
E_{Xv}}{kT_{\text{vib}}}\Bigr)}
\
\sum\limits_{N}
(2N+1)
\cdot
\exp{\Bigl(-\frac{\Delta E_{XvN}}{kT}\Bigr)}
}
=\mbox{[H$_{2}$]} \cdot
\frac{\exp{\Bigl(-\frac{\Delta
E_{Xv}}{kT_{\text{vib}}}\Bigr)}}{{\mathcal P}(T_{\text{vib}})},
\end{equation}
\narrowtext
where [H$_{2}$] -- total concentration of molecules, $\Delta 
E_{Xv}=E_{Xv0}-E_{X00}$, $\Delta E_{XvN}=E_{XvN}-E_{Xv0}$ -- vibrational and 
rotational energy differences, $T_{\text{vib}}$ and $T$ -- vibrational and 
translational temperatures, ${\mathcal P}(T_{\text{vib}})$ -- the vibrational 
partition function. So we assume that in all vibrational levels of the 
$X^{1}\Sigma_{g}^{+}$ state the populations of the rotational levels are in 
Boltzmann equilibrium with the gas temperature $T$.

Taking into account all the assumptions discussed above, the balance equation 
(\ref{eq:Balance}) may be written for the populations of $d^{3}\Pi_{u}^{-},v'$ 
vibronic states after summation over rotational levels in the following form
\begin{equation}
\frac{1}{{\mathcal P}(T_{\text{vib}})}
\sum\limits_v Q_{Xv}^{dv'} \cdot
\exp{\left(-\frac{\Delta
E_{Xv}}{kT_{\text{vib}}}\right)} \propto
\frac{N_{dv'}}{\tau_{dv'}}.
\label{eq:SlvTv}
\end{equation}

One may see that the determination of $T_{\text{vib}}$ may be achieved by 
numerical solution of the system of equations (\ref{eq:SlvTv})\footnote{It's 
also an inversion problem, but its solution is rather stable because it is 
quite easy to measure intensities of the first four bands ($v'=0-3$). That 
means at least three relative populations (\ref{eq:Ndv}), and therefore three 
equations (\ref{eq:SlvTv}) can be used for derivation of one unknown 
($T_{\text{vib}}$).} for several levels $v'$ with experimental data about 
$N_{dv'}$ derived from the intensities of the Fulcher-$\alpha$ Q-branch lines 
of diagonal ($v'=v''$) bands:
\begin{equation}
N_{dv'}=\sum\limits_N
\frac{I_{av'N}^{dv'N}}{A_{av'N}^{dv'N}}.
\label{eq:Ndv}
\end{equation}
This method of the $T_{\text{vib}}$ determination is based on sufficiently 
simple kinetics model and certain set of constants presented in Tables 
\ref{tab:FtA} and \ref{tab:FCF2}. It is even more simple then the predecessor 
\cite{LP85} because it needs only intensity measurements.

In the framework of our model the continuum intensity can be written as
\widetext
\begin{equation}
I_{ab}(\lambda,T_{\text{vib}}) \propto
\frac{1}{{\mathcal P}(T_{\text{vib}})}
\sum\limits_{v'}A_b^{av'}(\lambda)\cdot
\tau_{av'}\cdot
\sum\limits_{v=0}^{v'_{\text{max}}}Q_{Xv}^{av'}\cdot
\exp{\left(-\frac{\Delta E_{Xv}}{kT_{\text{vib}}}\right)
}.
\label{eq:Iab2}
\end{equation}
\narrowtext

To be able to work with relative intensities we may introduce the normalized 
relative intensity
\begin{equation}
J_{ab}(\lambda,\lambda_0,T_{\text{vib}}) =
I_{ab}(\lambda,T_{\text{vib}}) /
I_{ab}(\lambda_0,T_{\text{vib}}),
\label{eq:Inorm}
\end{equation}
which is equal to unity for $\lambda=\lambda_0$.

The normalized intensities (\ref{eq:Inorm}) may be tabulated for various 
values of $T_{\text{vib}}$ and then may be used for a determination of the 
vibrational temperature by least squares fitting (\ref{eq:chi2}) of 
experimental and calculated continuum intensity distributions. The results of 
our calculations for $\lambda_0=355.5$~nm and $T_{\text{vib}}$ = 0, 3000 and 
5000~K are shown in Fig.~\ref{fig:probab7}. One may see that a variation of 
vibrational temperature leads to certain change in the shape of the continuum 
intensity distribution. Most sensitive are the wavelength regions near 
$\lambda \approx 200$ and 300~nm. The first one is out of our range of 
observation and cannot be used for diagnostics herein. Nevertheless the fact 
should be considered as one of the results of our present work. It may be 
recommended for VUV spectroscopy if the problems of the sensitivity 
calibration in this difficult region would be solved somehow (with the use of 
synchrotron radiation, for example). The changes near $\lambda \approx 300$~nm 
are less pronounced, therefore they may be used in the following way.

When the vibrational temperature is sufficiently low and
\begin{equation}
Q_{X0}^{av'} \gg \sum\limits_{v=1}
Q_{Xv}^{av'}\cdot
e^{-\frac{\Delta E_{Xv}}{kT_{\text{vib}}}},
\end{equation}
then the excitation from vibronic levels with $v>0$ may be neglected and
\begin{equation}
J_{ab}(\lambda,\lambda_0,0) =
\frac{
\sum\limits_{v'=0}^{v'_{\text{max}}}
A_b^{av'}(\lambda)\cdot \tau_{av'}\cdot
Q_{X0}^{av'}
}{
\sum\limits_{v'=0}^{v'_{\text{max}}}
A_b^{av'}(\lambda_0)\cdot
\tau_{av'}\cdot Q_{X0}^{av'}
}
\label{eq:Il00}
\end{equation}
represents the shape of the continuum for $T_{\text{vib}}\to0$. It does not 
depend on discharge conditions and may be easily calculated because molecular 
constants in Eq.~(\ref{eq:Il00}) are known. The results of such calculation 
for $\lambda_0=355.5$~nm and $v'_{\text{max}}=6$ are presented in the last 
column of Table~\ref{tab:Alv}. One may see that in the wavelength range of 
observation, $\lambda = 225-400$~nm, the continuum intensity shows monotonic 
increase towards short wavelengths.

\begin{figure}
 \epsfxsize=8.6cm \epsfbox{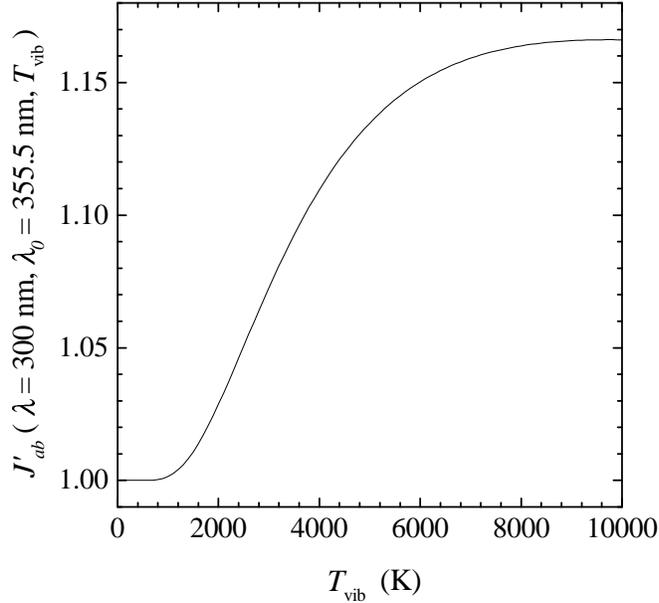}
 \caption{Normalized continuum intensity 
$J'_{ab}(\lambda,\lambda_0\mbox{=355.5~nm},T_{\text{vib}})$ calculated for 
$\lambda=300$~nm as a function of the vibrational temperature 
$T_{\text{vib}}$.}
 \label{fig:3}
\end{figure}

The ratio
\begin{equation}
J'_{ab}(\lambda,\lambda_0,T_{\text{vib}}) =
J_{ab}(\lambda,\lambda_0,T_{\text{vib}}) /
J_{ab}(\lambda,\lambda_0,0)
\label{eq:Ill0T}
\end{equation}
may also be calculated for various values of the vibrational temperature and 
then be used as a nomogram for the determination of $T_{\text{vib}}$ from 
measured relative continuum intensity without solving the inverse problem by a 
minimization of Eq.~(\ref{eq:chi2}). The results of such calculations for a 
limited number of $T_{\text{vib}}=0-6000$~K are shown in Fig.~\ref{fig:2}. One 
may see that the ratio (\ref{eq:Ill0T}) has some sensitivity to the 
vibrational temperature especially near 300~nm. For this particular wavelength 
we may organize a graph of 
$J'_{ab}(\lambda=300\mbox{~nm},\lambda_0=355.5\mbox{~nm},T_{\text{vib}})$ 
versus $T_{\text{vib}}$ presented in Fig.~\ref{fig:3} where the sensitivity is 
even more clear. This dependence may be used as a nomogram too, being 
especially useful for \textit{in situ} control of plasmachemical processes.

\section{Experimental setup}
\label{sec:experiment}

As it was already mentioned in the introduction the experiments in pure 
hydrogen and H$_{2}$+Ar plasmas were carried out in two different gas 
discharges.

The first one is a hot cathode arc with 2 mm diameter constriction analogous 
to those used in \cite{h2lamp1,h2lamp2}. The discharge device was filled with 
\mbox{8~mbar} of spectrally pure hydrogen. The range of the discharge current 
$I = 10-500$~mA allowed us to achieve rather high current densities $j \approx 
0.3-16$~A~cm$^{-2}$. 

The second plasma source was a H$_{2}$+Ar microwave discharge excited by a 
planar microwave applicator similar to that described in \cite{Ohl92b}. This 
plasma reactor has been specially designed for basic research on molecular 
microwave discharges mainly utilizing spectroscopic diagnostic methods 
\cite{RKL98}. The discharge configuration has the advantage of being well 
suited for end-on spectroscopic observations, because considerable homogeneity 
can be achieved to a certain extent, while side-on observations are also 
possible. Microwave plasma reactors of this planar type have been used for 
diamond deposition \cite{ORS93}, surface corrosion protection by deposition of 
organo-silicon compounds \cite{ROS93} and surface cleaning \cite{OSRKPS95}. 
The inner dimensions of the discharge vessel are 
15$\times$21$\times$120~cm$^3$ and the area of the quartz microwave windows 
was 8$\times$40~cm$^2$. The microwave power was fed in by a generator (SAIREM, 
GMP 12 KE/D, 2.45~GHz). We were able to measure the total input ($W = 1.5$~kW) 
and reflected power, the latter was in our conditions always negligible 
($<4$\%). The input power flux in the plane of the applicator windows was 
roughly estimated to about 4~W~cm$^{-2}$ assuming 20\% heat loss to coolant 
and environment. The plasma length in end-on direction was $15 \pm 3$~cm, 
mainly depending on the discharge pressure, and $8 \pm 1$~cm in side-on 
observation, while for absorption spectroscopy the end-on direction was 
preferred. Gas flow controllers and a butterfly valve in the gas exhaust allow 
to have an independent control of gas inflow ($\varphi$ = 100~sccm) and 
pressure ($p = 0.5-5$~mbar). The purity of gases was 99.996\%. H$_{2}$ and Ar 
are used as components of the feed gas mixture, since they are not only of 
interest for basic research but of importance for various applications as 
well. Our main interests were focused on studies of: (1) influence of Ar on 
H$_{2}$ emission, and (2) dissociation of molecular hydrogen.\footnote{
The role of atomic hydrogen is usually considered as very important for 
plasma-surface interaction processes \cite{RGGL94,LSKB96}, but it was not 
investigated in the planar microwave discharge.}

\begin{figure}
 \epsfxsize=8.6cm \epsfbox{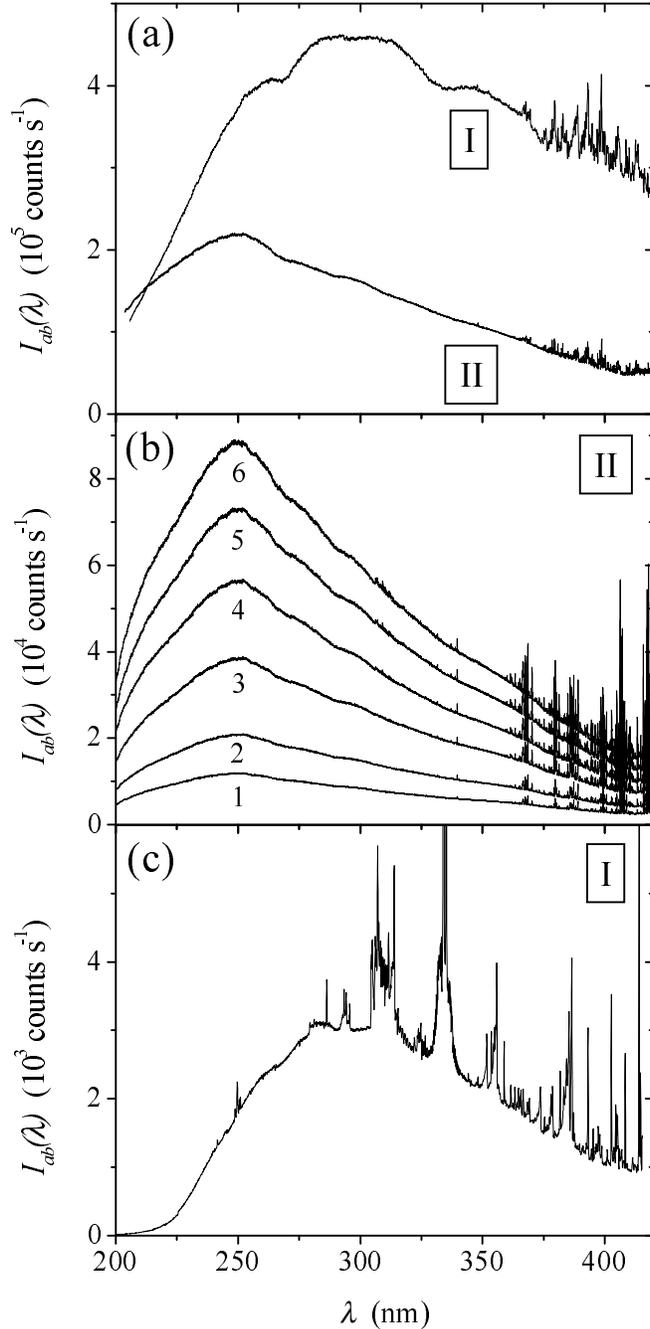}
 \caption{Experimental records (counts of CCD detector) of the continuum 
emission obtained in the dc capillary-arc discharges in (a) D$_{2}$, (b) 
H$_{2}$ with variation of the discharge current $i$ = 50, 100, 200, 300, 400, 
and 500~mA (curves 1, 2, 3, 4, 5, and 6) and (c) in the microwave discharge 
($p=0.5$~mbar, H$_{2}$:Ar=1:1). Spectral distribution of the sensitivity is 
not taken into account. The measurements have been performed with gratings 600 
(I) and 1200 (II) grooves per mm.}
 \label{fig:conti3}
\end{figure}

The central part of the plasma along the axis of the capillary discharge was 
focused by a quartz achromatic lens onto the entrance slit of a monochromator. 
In the microwave plasma the line of sight was 12~mm below the microwave 
windows of the applicator. For measurements of the continuum intensity we used 
a 0.5~m Czerny-Turner monochromator (Acton Research Corporation Spectra 
Pro--500) with gratings of 600 and 1200 grooves per mm in the first order. The 
line intensities of the Fulcher-$\alpha$ bands however were measured using a 
high resolution 1~m double monochromator of Czerny-Turner type with 1800 
grooves per mm (Jobin Yvon RAMANOR U1000) in the first order.\footnote{We 
started our measurements with the grating of 600 grooves per mm (blazed for 
300~nm) because of higher sensitivity [see Fig.~\ref{fig:conti3}(a)] important 
in the case of microwave discharge with its low intensities [pay attention to 
ordinate scales of Fig.~\ref{fig:conti3}(c) and \ref{fig:conti3}(a,b)]. Later 
we had to switch to the grating with 1200 grooves per mm (blazed for 250~nm) 
to get better relative sensitivity for shorter wavelengths [see 
Fig.~\ref{fig:conti3}(a)] and to select the continuum intensity from that of 
the band spectra of H$_{2}$ in more accurate way.} The output light was 
detected by a CCD matrix detector (TE-cooled, 512$\times$512 pixel, size of 
image zone: 9.7$\times$9.7~mm$^2$) of the Optical Multichannel Analyzer PARC 
OMA IV connected with a computer. Special software made it possible to collect 
and analyze the recorded spectra.

The wavelength calibration was done by means of known Hg, Ar and H$_{2}$ 
spectral lines. The relative spectral sensitivity of the spectroscopic system 
in the range $\lambda=225-450$~nm has been obtained using a deuterium lamp 
(Heraeus Noble Light DO651MJ) calibrated by the Physikalisch--Technische 
Bundesanstalt (Braunschweig). The absolute value of the sensitivity for 
$\lambda=350$~nm was determined with a tungsten ribbon lamp. Its 
current-temperature calibration has been performed by the Russian State 
Institute of Standards (St.-Petersburg). We used two different currents 
corresponding to the absolute temperatures of tungsten $T$ = 2637 and 2763~K. 
This gave us usually a spread in the values of the sensitivity (at 
$\lambda=350$~nm) of less than 1\%. The emissivity coefficient of the tungsten 
ribbon for these temperatures was taken from \cite{Malyshev79} and used for 
the calculation of the spectral distribution of emission intensity.

Some typical experimental records of the spectra of a standard deuterium lamp 
(a), pure hydrogen capillary-arc discharge (b) and H$_{2}$+Ar plasma of the 
microwave discharge (c) are presented in Fig.~\ref{fig:conti3}. The 
non-monotonic structure in the range $\lambda=250-350$~nm is connected with 
the spectral distribution of the sensitivity of our spectrometer. In the 
long-wavelength end of the first two records the band (actually multi-line) 
spectra of D$_{2}$ and H$_{2}$ molecules may be seen. One may see also that in 
spite of the gas flow mode our microwave plasma is not spectrally pure. It 
contains some impurities: oxygen (283~nm), OH-radical (bands at 283, 
$310-320$~nm), NH-radical (336~nm) and N$_{2}$ molecule (bands of the 2nd 
positive system $280-400$~nm mixed with H$_{2}$ bands). It should be noted 
that some lines are overexposed in the record shown in 
Fig.~\ref{fig:conti3}(c). The data of the continuum intensity, except those in 
Figs.~\ref{fig:probab7}, \ref{fig:2} and \ref{fig:conti3}, were manually 
extracted from the measured intensity distribution.

As can be seen from previous description, the spectroscopic part of our 
experimental setup is certain combination of components commercially available 
nowadays. Therefore it may be easily reproduced. Experimental errors are 
mainly caused not by the detection system but by the random noise of plasma 
sources used. Special attention should be paid to selection of the continuum 
intensity from that of multi-line spectrum of H$_{2}$. 

The results of our experiments will be presented together with corresponding 
analysis in the following sections. Perhaps most interesting of them are the 
first observations of:
(1) small changes in the shape of the continuum in pure hydrogen plasma caused 
by vibrational excitation in the ground electronic state of H$_{2}$ 
molecule,\footnote{That was achieved due to sufficient improvement of 
experimental technique in comparison with previous investigations 
\cite{Zaidel,LP85,LP88} where the changes have been masked by random noise of 
discharge and large systematic errors.}
(2) influence of Ar concentration on the shape of the continuum in H$_{2}$+Ar 
plasma of microwave discharge.

\section{Results and Discussion}
\label{sec:ResultsDiscussion}

\subsection{Pure hydrogen dc capillary-arc discharge}
\label{sec:dc}

\begin{figure}
 \epsfxsize=8.6cm \epsfbox{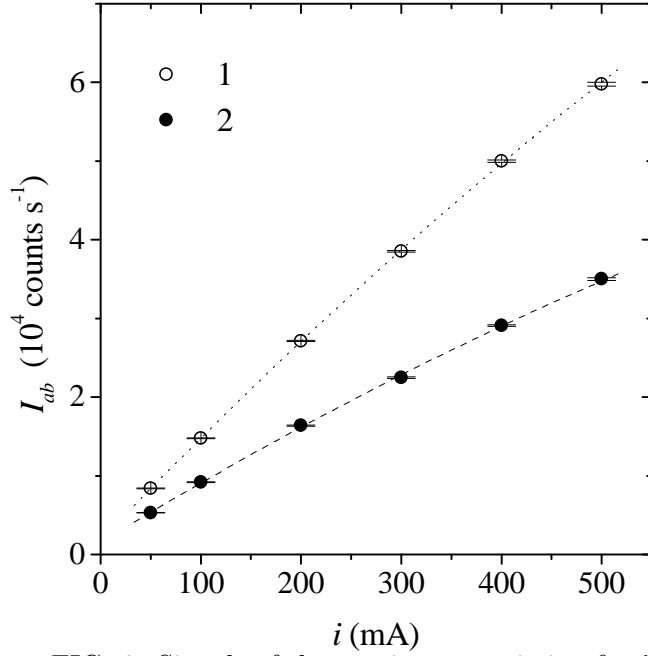}
 \caption{Signals of the continuum emission for $\lambda_0$ = 300 and 355.5~nm 
(points 1 and 2) against discharge current of H$_{2}$ capillary arc.}
 \label{fig:int}
\end{figure}

Relative intensities of the continuum and of the first Q-branch lines Q1 to Q5 
of Fulcher-$\alpha$ bands ($v'$=$v''$=0-5) were measured in the discharge 
current range mentioned above. The intensities were found to be proportional 
to discharge current if one takes into account a decrease of the density of 
molecules due to warming of the gas inside closed lamp by the hot cathode and 
discharge current. As an example the signals of CCD detector proportional to 
the continuum intensity are shown in Fig.~\ref{fig:int} as a function of the 
discharge current. The linearity observed may be considered as an argument in 
favor of direct electron impact excitation of the levels under the study.

\begin{figure}
 \epsfxsize=8.6cm \epsfbox{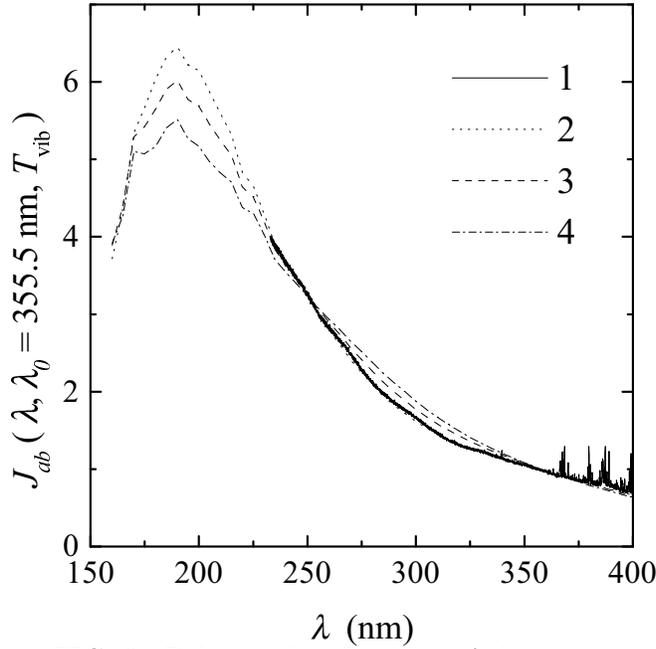}
 \caption{Relative distributions of the continuum intensity measured in the 
capillary-arc discharge with $I=100$~mA (curve 1) and those calculated for 
$T_{\text{vib}}$=0, 3000, 5000~K (curves 2, 3, and 4).}
 \label{fig:probab7}
\end{figure}

The shape of the continuum intensity distribution was observed to be 
independent on the discharge current for currents $i<100$~mA within 
experimental errors. A typical result used in further analysis is presented in 
Fig.~\ref{fig:probab7}. As one may see, measured continuum intensity is close 
to that calculated for $T_{\text{vib}}\to0$. The low wavelength margin is 
limited by a sharp decrease of the sensitivity of our spectrometric system. 
For wavelengths longer than 360~nm the overlap of the continuum with band 
spectra of H$_{2}$ becomes more and more pronounced.

The first question to answer is how many populations of upper vibronic states 
can be obtained from the intensity distribution measured in the wavelength 
range of observation. To find the optimum number of adjusted parameters 
$N_{av'}$ we made the $\chi^2$-minimization [Eq.~(\ref{eq:chi2})] with various 
$m=(v'_{\text{max}}+1)$. The results of the calculations, made for the 
experimental data of Fig.~\ref{fig:probab7}, are presented in 
Table~\ref{tab:RDVS1}. $\chi^2$ shows a monotonic decrease and becomes lower 
than unity for $v'_{\text{max}} \ge 3$. Another characteristic, presented in 
the table, is
\begin{eqnarray}
\rho_{\text{max}} = \max{(\rho_i)}\quad \mbox{with}\quad 
\rho_i=\sqrt{1-1/\Phi_i},\\
\mbox{where}\quad \Phi_i = Z_{ii} Z_{ii}^{-1}\quad\mbox{and}\quad 
Z_{lk}=\frac{\partial^2(\chi^2)}{\partial a_l \partial a_k}.
\end{eqnarray}
The magnitudes $\rho_i$ are characteristics of the precision of the solution. 
Its zero value corresponds to the limit case of total independence of the 
parameters, while $\rho_i=1$ means that the parameter $a_i$ may be expressed 
as a linear combination of the other parameters and therefore the number of 
adjusted parameters should be decreased. It may be seen from 
Table~\ref{tab:RDVS1} that in our case $\rho_{\text{max}}$ is growing with the 
increase of $m$; $\rho_{\text{max}}$ is close to unity for $v'_{\text{max}} 
>$3. The maximum value of the relative standard deviation of the parameters 
$(\Delta N_{av'}/N_{av'})_{\text{max}}$ (in our case it is usually observed 
for $v'$=$v'_{\text{max}}$) is small and almost independent on 
$v'_{\text{max}}$ for $v'_{\text{max}}=0-3$, then it jumps to about 0.83 at 
$v'=4$ and becomes much bigger for $v'=5,6$. It means that the standard 
deviation of $N_{a4}$ derived from our spectrum is almost the same as its 
optimal value.

Thus in our case a determination of the populations of more than four vibronic 
levels is meaningless. The reason is very simple and may be seen from 
Table~\ref{tab:RDVS2} where the relative populations $N_{av'}/N_{a0}$ 
(obtained for $v'_{\text{max}}=3$) are shown together with relative 
intensities of various $a^{3}\Sigma_{g}^{+},v'\to b^{3}\Sigma_{u}^{+}$ 
transitions in the spectral range under the study ($\lambda=225-400$~nm)
\begin{equation}
\xi_{v'}=
\frac{
\int\nolimits_{\lambda_1}^{\lambda_2}N_{av'}\cdot
A_{b}^{av'}(\lambda)\,d\lambda
}{
\int\nolimits_{\lambda_1}^{\lambda_2}I_{ab}(\lambda)\,d\lambda
}.
\label{eq:xi}
\end{equation}
From the data of Table~\ref{tab:RDVS2} it becomes clear that only 4\% of the 
detected emission quanta originate from levels with $v'>3$. The populations 
$N_{av'}$ observed in the pure hydrogen capillary-arc discharge for currents 
$i=10-100$~mA are quite close to those calculated in the framework of our 
simple model with $T_{\text{vib}}\to0$. It means that the populations of the 
$X^{1}\Sigma_{g}^{+},v$ vibronic levels with $v>0$ are so small that they do 
not affect relative populations of $a^{3}\Sigma_{g}^{+},v'$ levels and 
therefore relative continuum intensity distribution.

\begin{figure}
 \epsfxsize=8.6cm \epsfbox{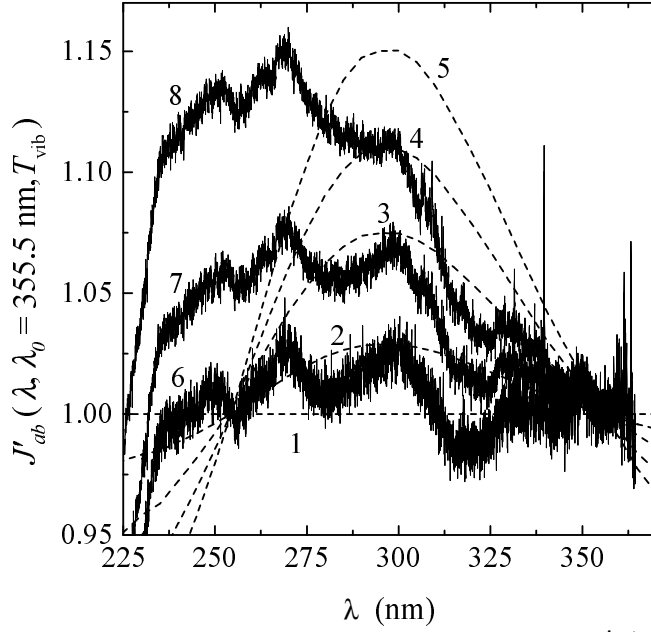}
 \caption{Normalized continuum intensity 
$J'_{ab}(\lambda,\lambda_0\mbox{=355.5$\,$nm},T_{\text{vib}})$ calculated by 
Eq.~\protect(\ref{eq:Ill0T}) for various vibrational temperatures 
$T_{\text{vib}}$ = 0, 2000, 3000, 4000, 6000~K (curves 1, 2, 3, 4, and 5) and 
measured in capillary arc discharge at currents $i$ = 50, 200, and 300~mA 
(curves 6, 7, and 8).}
 \label{fig:2}
\end{figure}

\begin{figure}
 \epsfxsize=8.6cm \epsfbox{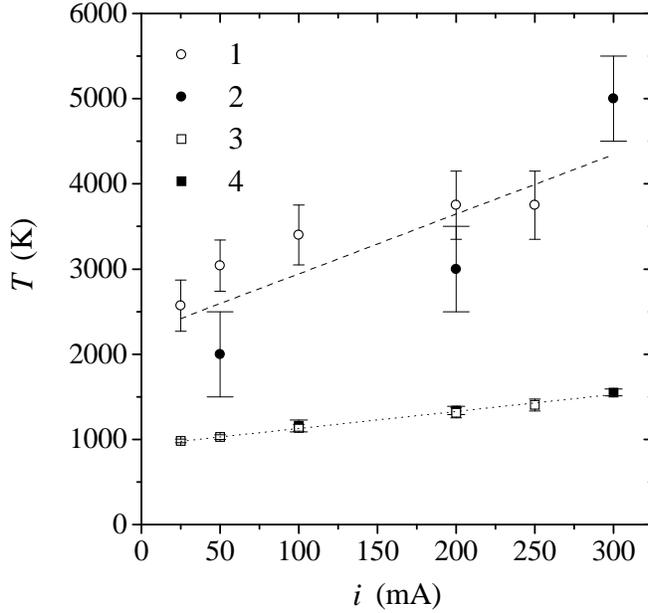}
 \caption{Vibrational (1,2) and rotational (3,4) temperatures in the ground 
$X^{1}\Sigma_{g}^{+}$ state of the H$_{2}$ in the capillary-arc pure hydrogen 
plasma for various discharge currents $i$ determined from the continuum 
intensity distribution (1) and from the emission of Fulcher-$\alpha$ bands 
(2-4). The data (1-3) -- present work, (4) -- Ref.~\protect\cite{LT82}.}
 \label{fig:risynok5}
\end{figure}

For currents higher than $50-100$~mA we observed some systematic changes in 
the shape of the continuum intensity distributions. As an example our 
experimental data on the normalized intensities 
$J'_{ab}(300,355.5,T_{\text{vib}})$ from Eq.~(\ref{eq:Ill0T}) are presented 
for three values of the discharge current in Fig.~\ref{fig:2} together with 
the curves calculated with our model for various values of $T_{\text{vib}}$. 
One may see that two main predictions of our model - the maxima near 300~nm 
and crossing with $J_{ab}'=1$ near 250~nm - are in qualitatively good 
accordance with the experiment. The experimental curves are a bit shifted to 
shorter wavelengths. To treat the data quantitatively one should take into 
account that the differences between experimental and theoretical curves are 
comparable with errors of both the experiment and modeling:
\begin{enumerate}
\item[(1)] The actual random noise of our measurements is about two times 
bigger than that shown in Fig.~\ref{fig:2} due to smoothing during calibration 
procedure. It was made for better visibility of the trend.
\item[(2)] The normalized intensity $J_{ab}'$ is the ratio of the measured 
intensities, so its errors are twice more than that of measured intensities. 
\item[(3)] The systematic errors of the measurement may be even bigger because 
errors of the sensitivity calibration should be included, while the errors of 
the relative energy calibration of the deuterium lamp are about 4\%. 
\item[(4)] Errors of the calculations are not so easy to estimate, but most 
likely they may be of the same order of magnitude. They are caused by the 
approximations made in the formulation of our very simple 
excitation-deactivation model [Eq.~(\ref{eq:Balance})] and by the 
uncertainties of the rate coefficients (see Section~\ref{sec:Constants}).
\end{enumerate}
Nevertheless, we used the values of maximum deviation of experimental curves 
from  $J_{ab}'=1$ level for estimation of $T_{\text{vib}}$. The results are 
shown in Fig.~\ref{fig:risynok5} together with values of $T_{\text{vib}}$ 
obtained from Fulcher-$\alpha$ band intensities by a system of 
Eqs.~(\ref{eq:SlvTv}) for $v'=0-3$.\footnote{Intensities of first five lines 
Q1-Q5 were used in each band.} One may see that both spectroscopic methods of 
$T_{\text{vib}}$ determination developed in the present work are in good 
accordance. 

The ground state rotational temperature was obtained from the intensity 
distributions in Fulcher-$\alpha$ Q branches by the method proposed in 
\cite{L80}. In our conditions the rotational and translational (gas) 
temperature are known to be in rather good accordance 
\cite{FrontII1,L84,AKKKLOR96}. The results are also shown in 
Fig.~\ref{fig:risynok5} together with earlier data obtained in \cite{LT82}. 
One may see that our measurements show remarkable difference between 
$T_{\text{vib}}$ and $T$ analogous to that observed in other conditions by 
CARS \cite{Capitelli86}.

For currents higher than 300~mA the continuum intensity distribution became 
again insensitive to the current variation. This is most probably caused by 
large values of gas temperature and concentration of atomic hydrogen that is 
favorable for the acceleration of the VT-relaxation. On the other hand this 
behavior can also be explained by our model, since there is a certain upper 
limit above which the continuum intensity distribution is no longer sensitive 
to an increase of vibrational temperature (see Fig.~\ref{fig:3}).

\subsection{Microwave discharge in H$_{2}$+Ar mixture}
\label{sec:mw}

\begin{figure}
 \epsfxsize=8.6cm \epsfbox{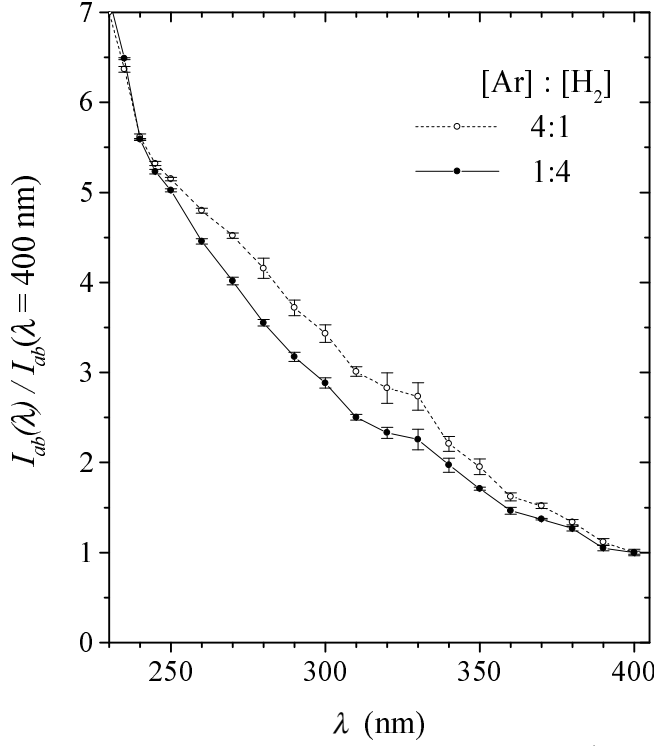}
 \caption{The relative continuum intensity (normalized for unity at 
$\lambda=400$~nm) measured in a H$_{2}$+Ar microwave plasma for two different 
ratios of the components at $p=0.5$~mbar.}
 \label{fig:mw1a}
\end{figure}

The results of our measurements show that the spectral intensity distribution 
of the continuum in the H$_{2}$+Ar microwave discharge has a noticeably 
different shape compared to that in pure hydrogen. Moreover, the shape was 
found to depend on the ratio of the components in the mixture 
([Ar]:[H$_{2}$]), which was varied from (19:1) to (1:4) under the same total 
pressure and power input. 
We observed a relative increase of the continuum emission in the wavelength 
region near 300~nm. As an illustration some typical results for (4:1) and 
(1:4) mixtures are shown in Fig.~\ref{fig:mw1a}. One may see that the 
additional emission is located in the same wavelength interval as 
$a^{3}\Sigma_{g}^{+},v'$=0$\to b^{3}\Sigma_{u}^{+}$ transition of H$_{2}$ (see 
Fig.~\ref{fig:1}) and ArH* (see below). Note that measurements in between 
exhibit a monotonic dependence of the increased emission in the wavelength 
range around 300~nm. An analysis of various possible explanations of the 
observed phenomenon lead us to the assumption that in our conditions an 
excitation transfer from excited Ar atoms to the hydrogen molecules might take 
place.

This experimental observation leads to two important consequences: 
\begin{enumerate}
\item[(1)] The shape of the continuum in H$_{2}$+Ar plasma is affected by two 
different excitation processes, therefore our model based on one of them 
(direct electron impact excitation) can not be used for the determination of 
$T_{\text{vib}}$.\footnote{The further analysis shows that this loss is well 
compensated by the opportunity to get information about the mechanism of 
excitation of the continuum emission and to account for the excitation 
transfer in determination of the total dissociation rate.} 
\item[(2)] To check the hypothesis of excitation transfer in our plasma we 
have to accompany our emission intensities with absorption measurements of 
populations of long-living states of Argon atoms (see Section 
\ref{sec:ArLevel}).
\end{enumerate}
\subsubsection{Ar*$\to$H$_{2}$ excitation transfer, prehistory}

Excitation transfer in collisions of metastable ($^{3}$P$^{\circ}_{0,2}$) and 
resonant ($^{3}$P$^{\circ}_{1}$ and $^{1}$P$^{\circ}_{1}$) Ar atoms with 
hydrogen molecules is known since the pioneer work of Lyman \cite{Lyman11} and 
was investigated so far in plasma and beam experiments mainly as: (1) 
quenching of excited Ar and (2) additional excitation of VUV Lyman and Werner 
bands ($B^{1}\Sigma_{u}^{+}$, $C^{1}\Pi_{u}$ $\to$ $X^{1}\Sigma_{g}^{+}$) of 
H$_{2}$. The first attempts to study the role of this effect in the emission 
of the hydrogen dissociation continuum in H$_{2}$+Ar plasmas appeared quite 
recently \cite{FrontII1,DissM96,LMT97,LM98}.

The excitation of $a^{3}\Sigma_{g}^{+}$ state of H$_{2}$ due to the excitation 
transfer from excited Ar* atoms to the ground state H$_{2}$ molecule may go in 
two different ways:
\widetext
\begin{equation}
\mbox{Ar$^*$(4s$^{1,3}$P$^{\circ}_{J}$) + H$_{2}$($X^{1}\Sigma_{g}^{+} ,v,N$) 
+ $T_0$ \hspace{5mm} $\to$ \hspace{5mm} (ArH$_{2}$)$^*$} \hspace{5mm} 
\to\label{eq:R2}
\end{equation}
$$
\hspace*{2mm}\to\left\{
\begin{array}{c}
\mbox{H$_{2}$*($a^{3}\Sigma_{g}^{+},v',N'$) + Ar($^{1}$S$_0$) $\to$ 2 
H($1^{2}$S$_{1/2}$) + Ar($^{1}$S$_0$)} + T_1 + 
h\nu(\mbox{H$_{2}$})\hspace{18mm} (\ref{eq:R2}a)
\\[12pt]
\mbox{ArH*($A^{2}\Sigma,v',N'$) + H($1^{2}$S$_{1/2}$) $\to$ 2 
H($1^{2}$S$_{1/2}$) + Ar($^{1}$S$_0$)} + T_2 + h\nu(\mbox{ArH}).\hspace{11mm} 
(\ref{eq:R2}b)
\end{array}
\right.
$$
\narrowtext
Here (ArH$_{2}$)* denotes a temporary excited state of the system during the 
interaction, $T_0$ and $T_1$, $T_2$ are the initial and final kinetic energies 
of reacting particles in the mass center frame. The branching ratio between 
these two output channels of the reaction should depend on the initial excited 
state of the Ar atom and the vibro-rotational state of the molecule as well as 
on the collision energy $T_0$. As far as we know it is not established up to 
now in spite of a certain attempt in crossed beam experiments \cite{LFSM85}.

It may be shown, that if one wants to consider the excitation transfer 
possibility for $^{1,3}$P$^{\circ}_{J}$ argon atoms with various $J$ values 
from the energy point of view, the existence of vibro-rotational and 
translational energies of H$_{2}$ molecules in plasma should be taken into 
account \cite{LM98}. Then in our conditions the observed effects are most 
likely due to the collisions of the resonant $^{3}$P$^{\circ}_{1}$ and 
metastable $^{3}$P$^{\circ}_{2}$ argon atoms with the 
$X^{1}\Sigma_{g}^{+},v$=0 hydrogen molecules.

\subsubsection{3s$^2$3p$^5$4s level populations of Ar \label{sec:ArLevel}}

To clear up the situation we measured the population densities of the Ar 
3s$^2$3p$^5$4s ($^{3}$P$^{\circ}_{0,1,2}$ and $^{1}$P$^{\circ}_{1}$) levels by 
an ordinary self-absorption method with one mirror behind the discharge vessel 
\cite{Zebra}.

Neglecting Stark broadening of the spectral lines the population of the 
initial state of $k \leftarrow i$ absorption transition may be presented as
\begin{equation}
N_i = \sqrt{\frac{\pi}{4\ln{2}}}\ \frac{\Delta\nu_D}{f_0 f_{ik}}\ \kappa_0,
\end{equation}
where
\begin{equation}
f_0 = \frac{\pi e^2}{m_e c} = 2.64 \cdot 10^{-2}\ \mbox{cm}^2\ \mbox{s}^{-1},
\end{equation}
$f_{ik}$ -- the oscillator strength for the $k \leftarrow i$ transition in 
units of $f_0$, $\Delta\nu_D$ -- the Doppler half-width depending on the gas 
temperature.

The absorption coefficient $\kappa_0$ was obtained by numerical solutions of 
the following non-linear equation \cite{Zebra}:
\begin{equation}
2 \left[1-\frac{S(2\kappa_0 l)}{S(\kappa_0 l)}\right] = 
\frac{(1+r)-I_{m}/I_0}{r},
\end{equation}
where $I_{m}$ and $I_0$ -- the line intensities measured with and without the 
mirror, $l$ -- length of plasma column along the axis of observation, $r$ -- 
the reflection coefficient of the mirror and $S(x)$ -- the Ladenburgh-L\'evy 
function calculated by formulas from \cite{Frisch63}.

The effective reflection coefficient has been determined experimentally by the 
measurements of several Ar lines (603.1, 703.0, 714.7, 789.1, 860.6, 862.0, 
and 876.1~nm) which are free of self-absorption. It was found to be $r=0.52\pm 
0.02$ and independent on $\lambda$. We used the values of the oscillator 
strengths from \cite{Radzig}.

The intensities of two or three different spectral lines were used for the 
population density determination of every 3s$^2$3p$^5$4s level, namely:\\
\begin{enumerate}
\item[$^{3}$P$^{\circ}_{2}$] -- 763.511, 801.479, 811.531~nm
\item[$^{3}$P$^{\circ}_{1}$] -- 751.465, 738.398, 810.369~nm
\item[$^{3}$P$^{\circ}_{0}$] -- 794.815, 866.794~nm
\item[$^{1}$P$^{\circ}_{1}$] -- 826.452, 840.821~nm.
\end{enumerate}
The conditions were the same as those used in our measurements of the 
continuum intensity.

\begin{figure}
 \epsfxsize=8.6cm \epsfbox{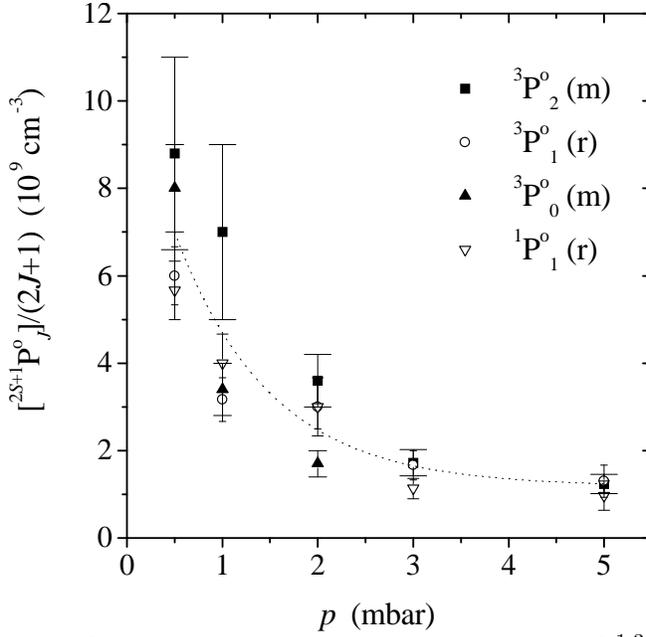}
 \caption{The scaled population densities of $^{1,3}$P$^{\circ}_{J}$ levels of 
Ar as a function of total pressure $p$ for constant gas mixture H$_{2}$+Ar 
(1:1) in the microwave discharge plasma. The total average population of all 
3s$^2$3p$^5$4s levels may be obtained by multiplication of 12 to the dashed 
curve.}
 \label{fig:mw3a}
\end{figure}

Typical results are shown in Fig.~\ref{fig:mw3a}. The scaled populations are 
absolute values of the level population densities divided by their statistical 
weights $(2J+1)$. They have the meaning of the average populations of the 
multiplet structure sublevels. In equilibrium conditions they should be almost 
equal to each other because the levels have almost the same energy -- 11.55, 
11.62, 11.72, and 11.83 eV for $^{3}$P$^{\circ}_{2}$, $^{3}$P$^{\circ}_{1}$, 
$^{3}$P$^{\circ}_{0}$, and $^{1}$P$^{\circ}_{1}$. One may see from 
Fig.~\ref{fig:mw3a} that in our conditions:
\begin{enumerate}
\item
  The populations of various sublevels of all resonant and metastable states 
coincide within the experimental errors. It is a direct manifestation of long 
effective lifetimes of the resonant levels (due to self-absorption of the 
resonant radiation) and a high enough rate of collisional transitions between 
the 3s$^2$3p$^5$4s levels. So the metastable and resonant levels of Ar are 
actually mixed in the plasma under the study.
\item
  The sublevel populations show the monotonic decrease with total pressure 
which may be connected with: a decrease of microwave power applied to plasma 
in the volume of observation, a decrease in the rate coefficient of electron 
impact excitation (due to certain changes of electron velocity distribution) 
and an increase of collisional quenching.
\item
  The total population of all 12 sublevels may reach $10^{11}$ cm$^{-3}$ at 
low pressures. The rate coefficients of the Ar*$\to$H$_{2}$ excitation 
transfer are approximately $10^{-10}$~cm$^3$~s$^{-1}$ \cite{GP80}. Thus the 
rate of the continuum excitation in the reaction is about 
10$^{16}$~cm$^{-3}$~s$^{-1}$. This is comparable with the observed total 
continuum emission (see Table~\ref{tab:RMW}) and should be taken into account 
together with direct electron impact excitation of the $a^{3}\Sigma_{g}^{+}$ 
state of H$_{2}$.
\end{enumerate}

\subsubsection{Shape of the continuum in H$_{2}$+Ar plasma}

Let us assume the spectral distribution of the continuum intensity as the sum 
of three independent terms corresponding to three different excitation 
mechanisms:
\begin{equation}
I_{ab}^{\text{calc}}(\lambda) = A [
c_1 I_{ab}(\lambda,0) +
c_2 I_b^{a0}(\lambda) +
c_3 I_{\text{ArH}}(\lambda)
],
\label{eq:I3}
\end{equation}
where $A$ -- absolute scale normalization constant equal to total continuum 
intensity in the range of observation $\lambda=225-400$~nm,  
$I_{ab}(\lambda,0)$ -- spectral distribution of the continuum radiation caused 
by the electron impact excitation of H$_{2}$ from lowest 
$X^{1}\Sigma_{g}^{+},v$=0 vibronic ground state [Eq.~(\ref{eq:Il00})],  
$I_b^{a0}(\lambda)$ -- the same  due to the reaction (\ref{eq:R2}a) for $v'=0$ 
only, $I_{\text{ArH}}(\lambda)$ -- the spectral distribution corresponding to 
the reaction (\ref{eq:R2}b) calculated in \cite{LFSM85}, $c_i$ -- the 
coefficients representing relative contributions of the components ($\sum_i 
c_i=1$). The spectral distributions  $I_{ab}(\lambda,0)$, $I_b^{a0}(\lambda)$ 
and $I_{\text{ArH}}(\lambda)$ where normalized for unit area in the range of 
wavelengths $\lambda=225-400$~nm.

\begin{figure}
 \epsfxsize=8.6cm \epsfbox{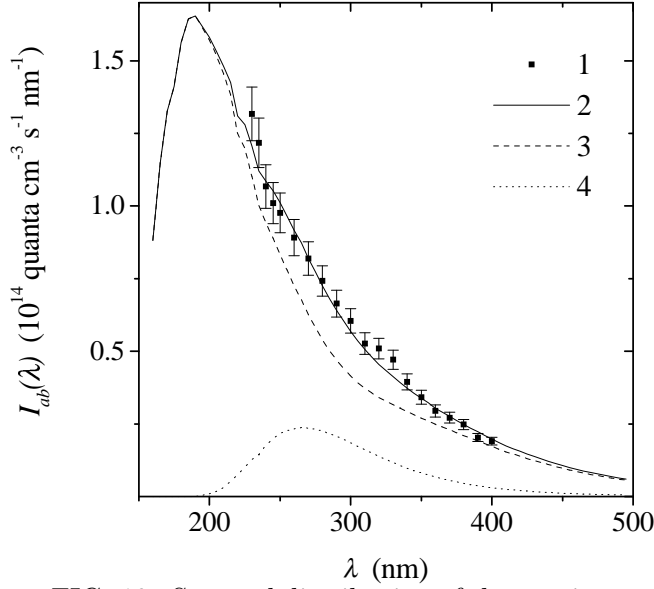}
 \caption{Spectral distribution of the continuum emission in the H$_{2}$+Ar 
(1:1) microwave plasma for the total pressure $p$ = 0.5 mbar. 1 -- 
experimental data points,  2 -- calculated by Eq.~\protect(\ref{eq:I3}) with 
the optimal set of parameters $A,c_1,c_2,c_3$ shown in 
Table~\protect\ref{tab:RMW}, 3 -- the contribution of the electron impact 
excitation for $T_{\text{vib}}\to0$ with $v'_{\text{max}}$=3, 4 -- 
contribution of the Ar*$\to$H$_{2}$ excitation transfer to 
$a^{3}\Sigma_{g}^{+},v'$=0.}
 \label{fig:mw2a}
\end{figure}

Then the spectral distributions of the continuum measured in the microwave 
plasma under various conditions were fitted by the function (\ref{eq:I3}) 
according to Eq.~(\ref{eq:chi2}).  Typical results are presented in 
Table~\ref{tab:RMW} for five selected mixtures of Ar+H$_{2}$ from (1:4) up to 
(19:1). The first set of $\chi^2$-minimizations marked as ($c_1,c_2,c_3 \neq 
0$) was made without any additional preconditions. One may see that the 
observed shape of the continuum may be described by three adjusted parameters 
of the expansion (\ref{eq:I3}) with rather good precision ($\chi^2=1-2$). 
Moreover the fittings show the negligible contribution of the reaction 
(\ref{eq:R2}b), which cannot be considered as significant because the values 
of $c_3$ are comparable with standard deviations of their determination. 
Figure~\ref{fig:mw2a} illustrates typical results of such fitting. The 
contribution of the reaction~(\ref{eq:R2}b) is too small to be visible in the 
figure.

The second set of fittings has been performed under the additional condition 
$c_3=0$. This version of the model provides almost the same precision in the 
description of the observed spectra taking into account only the first two 
mechanisms included into Eq.~(\ref{eq:I3}).

In contrast the third set of fittings has been carried out for $c_2=0$ thus 
neglecting a contribution of the reaction (\ref{eq:R2}a). Again we got almost 
the same quality of fitting. These results are the direct consequence of the 
similarity of spectral distributions of the continuum emission caused by the 
reactions (\ref{eq:R2}a)  for $v'=0$ and (\ref{eq:R2}b).

It should be noted, that \textit{ab initio} potential curves of ArH* where 
never checked experimentally and the spectral distribution of the ArH 
continuum has been calculated in \cite{LFSM85} in the rough approximation of 
harmonic oscillator. The precision of such calculations cannot be high 
nowadays. Taking this into account we may assume that the insignificant 
contribution of the reaction  (\ref{eq:R2}b) obtained in the first set of 
fittings is an accidental numerical result. Most probably it is without any 
physical meaning because two other sets [based on directly opposite 
assumptions about the branching ratio between reactions (\ref{eq:R2}a) and 
(\ref{eq:R2}b)] gave us almost the same $\chi^2$ values.

Therefore, we have to conclude that the precision and the wavelength range of 
our measurements together with the lack of dependable ArH* transition 
probabilities do not allow us to distinguish contributions of the reactions 
(\ref{eq:R2}a) and (\ref{eq:R2}b). On the other hand, there is no doubt that 
in our conditions the role of the Ar*$\to$H$_{2}$ excitation transfer is well 
noticeable (10-30\%) both in excitation and dissociation rates (see below).

\subsection{Estimation of the rate of electron impact dissociation
\label{sec:dissrate}
}

As it was already mentioned in the introduction, the absolute value of the 
intensity of the continuum is directly connected with the rate of the 
radiative dissociation via spontaneous emission to the repulsive 
$b^{3}\Sigma_{u}^{+}$ state. If self-absorption in the plasma volume may be 
neglected, then the rate of radiative dissociation due to $a^{3}\Sigma_{g}^{+} 
\to b^{3}\Sigma_{u}^{+}$ transitions is equal to the total continuum intensity 
(expressed in number of photons per cm$^3$ in a second) integrated over the 
entire wavelength range
\begin{equation}
\left(\frac{d[\mbox{H$_{2}$}]}{dt}\right)_{ab} =
\int\nolimits_0^\infty I_{ab}(\lambda) d\lambda = 
\sum\limits_{v'=0}^{v'_{\text{max}}}
\frac{N_{av'}}{\tau_{av'}}.
\label{eq:dissrate}
\end{equation}

The wavelength range where we were able to measure the intensity was limited 
(see Fig.~\ref{fig:probab7}, \ref{fig:conti3}, \ref{fig:mw1a}). So we detected 
only a part of the continuum emission.\footnote{Measurements in VUV are 
possible in principle, but they are not typical in most gas discharge and 
plasma technology applications due to their technical complexity and the 
difficulties in the sensitivity calibration and the spatial resolution.} The 
rest should be obtained by some extrapolation. In pure hydrogen discharge 
plasma the vibrational temperature is usually rather low and the 
Eqs.~(\ref{eq:Iab}) and (\ref{eq:Inorm}) may be used for the extrapolation. 
The calculations show that in our case: (1) The dependence of the total 
intensity (area under the curve) on the $v'_{\text{max}}$ is insignificant 
after $v'_{\text{max}}$ became high enough. (2) The contribution of the 
undetected part of the spectrum is rather significant (about 60\% in the 
example shown in Fig.~\ref{fig:2}).

The other, more general way of treating of spectroscopic data is a numerical 
solution of the reverse problem by $\chi^2$-minimization (\ref{eq:chi2}). The 
relative or absolute populations $N_{av'}$ thus obtained may be used for the 
determination of the relative or absolute radiative dissociation rate. 
Radiative lifetimes are known, so an integration over wavelengths is not 
necessary and the procedure is reduced to only few summations in 
Eq.~(\ref{eq:dissrate}).

In low-pressure gas discharges the upper levels of the continuum transition 
are predominantly populated by $a^{3}\Sigma_{g}^{+} \gets X^{1}\Sigma_{g}^{+}$ 
electron impact excitation \cite{LP85,LP88}. A dissociation of hydrogen 
molecules is also mainly caused by electron impact excitation of various 
intermediate excited electronic states, bound and repulsive ones. The relative 
contribution of various channels to the total electron impact dissociation 
cross section $\sigma_{\text{diss}}(\varepsilon)$ was analyzed in 
\cite{CLL75}, later supplemented with an application of the close-coupling 
calculations of the cross-sections \cite{CL78}. It was found that 
singlet-triplet transitions play the major role, $b^{3}\Sigma_{u}^{+} \gets 
X^{1}\Sigma_{g}^{+}$ and $a^{3}\Sigma_{g}^{+} \gets X^{1}\Sigma_{g}^{+}$ being 
most prominent. The experimental data for $\sigma_{\text{diss}}(\varepsilon)$ 
are available from \cite{Corrigan65}.

All these cross-sections seem to have a rather similar shape as a function of 
the collision energy $\varepsilon$, except in the small region near the 
threshold, because an excitation of various electronic states obviously has 
different start-up energies. In \cite{FrontII1} we proposed to use average 
values of ratios between the cross sections (determined far enough from 
threshold region) for the estimation of the rate of the radiationless 
$b^{3}\Sigma_{u}^{+} \gets X^{1}\Sigma_{g}^{+}$ process 
$(d[\mbox{H$_{2}$}]/dt)_{b\gets X}$ and of the total dissociation rate 
$(d[\mbox{H$_{2}$}]/dt)_{\text{diss}}$ by the multiplication of the measured 
total continuum intensity with certain coefficients (4 and 15 
correspondingly). Actually, the coefficients depend on the shape of the 
electron energy distribution function and the total electron impact 
dissociation rate
\begin{equation}
\left(\frac{d[\mbox{H$_{2}$}]}{dt}\right)_{\text{diss}} =
\frac{\alpha_{\text{diss}}}{\alpha_{a\gets X}} 
\left(\frac{d[\mbox{H$_{2}$}]}{dt}\right)_{ab}.
\label{eq:dissrate2}
\end{equation}
Generally, the rate coefficients in (\ref{eq:dissrate2}) should be calculated 
with actual $F(\varepsilon)$ by Eq.~(\ref{eq:alpha}) \cite{LP85,LP88}. If the 
distribution function is not available like in our present study, only rough 
estimations are possible. It is quite obvious that, when 
$\varepsilon$-dependences of the cross-sections have similar shapes, the ratio 
of the rate coefficients is less sensitive to the near-threshold region for 
$F(\varepsilon)$ with a high content of fast enough electrons. To check the 
tendency we calculated the ratios for a Maxwellian distribution with an 
electron temperature of $T_e=1-10$~eV and the cross sections from 
\cite{CL78,Corrigan65}. The results are presented in Fig.~\ref{fig:rc}. One 
may see that both ratios show sharp decrease for $T_e<5$~eV becoming almost 
independent from $T_e$ for $T_e>5$~eV. It is quite typical for a low-pressure 
hydrogen discharge plasma to have a strongly non-Maxwellian $F(\varepsilon)$ 
with a high density of fast electrons having energies sufficiently higher than 
the thresholds (see i.e.~\cite{BLP87}). So we may use the asymptotes of 
Fig.~\ref{fig:rc} for the estimations. Numerical values of the coefficients are
\begin{equation}
\frac{\alpha_{b\gets X}}{\alpha_{a\gets X}} \sim 4.9\qquad\mbox{and}\qquad
\frac{\alpha_{\text{diss}}}{\alpha_{a\gets X}} \sim 16 .
\label{eq:aovera}
\end{equation}

In the pure hydrogen capillary-arc discharge we were not able to carry out 
absolute measurements because the plasma is inhomogeneous along the axis of 
observation and the effective length of the emitting plasma column is 
uncertain \cite{BLP87}.

\begin{figure}
 \epsfxsize=8.6cm \epsfbox{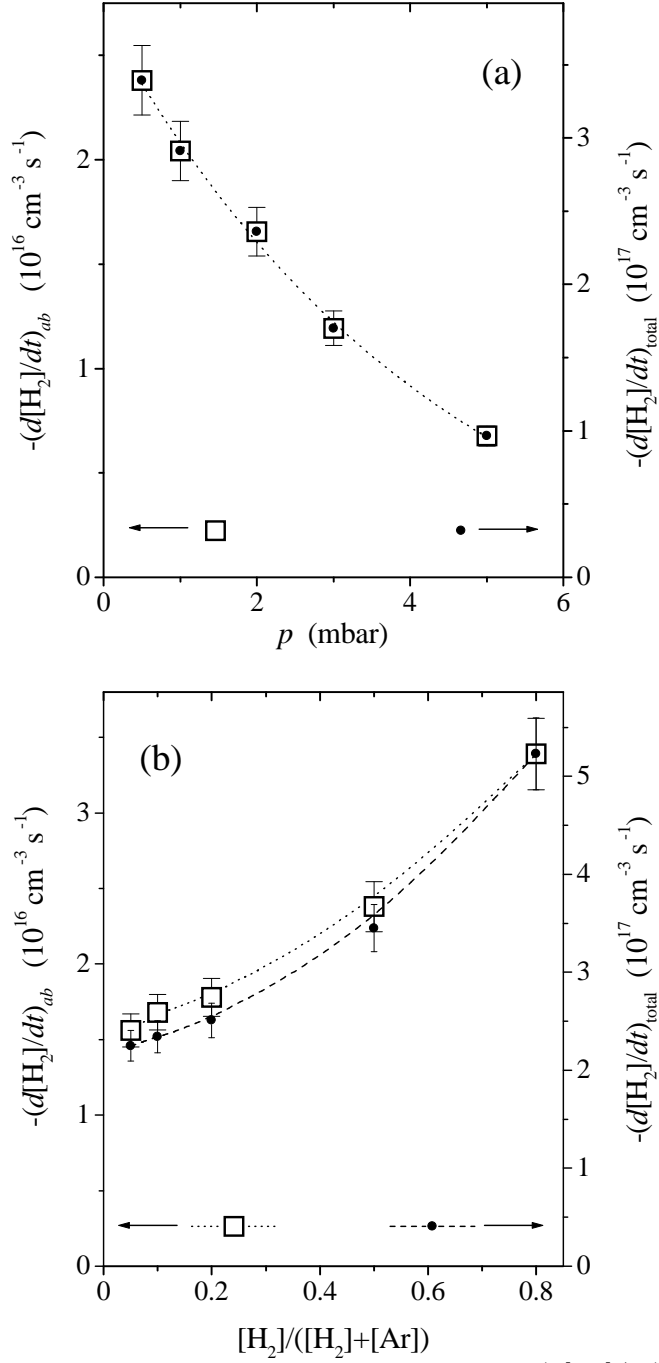}
 \caption{Radiative dissociation rate $(d[\mbox{H$_{2}$}]/dt)_{ab}$ ($\Box$) 
and total electron impact dissociation rate 
$(d[\mbox{H$_{2}$}]/dt)_{\text{total}}$ ($\bullet$) obtained in the H$_{2}$+Ar 
microwave discharge: (a) as a function of the total pressure $p$ for constant 
gas mixture (1:1) and (b) for various gas mixtures under constant total 
pressure $p$ = 0.5 mbar.}
 \label{fig:RMW}
\end{figure}

However, in the H$_{2}$+Ar microwave plasma the length of emitting column 
($8\pm1$)~cm is well defined and absolute continuum intensities were measured 
in a certain range of discharge conditions. But the Eqs.~(\ref{eq:dissrate}) 
and (\ref{eq:aovera}) are not valid any more, because Ar*$\to$H$_{2}$ 
excitation transfer plays an important role. The continuum intensity may be 
expressed as (\ref{eq:I3}) and the total dissociation rate is a sum of the 
terms corresponding to the electron impact and the excitation transfer. It can 
be shown that in this case the total dissociation rate
\widetext
\begin{eqnarray}
\left(\frac{d[\mbox{H$_{2}$}]}{dt}\right)_{\text{total}} & = &
\left(\frac{d[\mbox{H$_{2}$}]}{dt}\right)_{\text{diss}} +
\left(\frac{d[\mbox{H$_{2}$}]}{dt}\right)_{\text{Ar*}} \\
& = & A \left[
\frac{\alpha_{\text{diss}}}{\alpha_{a\gets X}}\,
c_1 \int\nolimits_0^\infty I_{ab}(\lambda,0)\, d\lambda +
c_2 \int\nolimits_0^\infty I_b^{a0}(\lambda)\, d\lambda +
c_3 \int\nolimits_0^\infty I_{\text{ArH}}(\lambda)\, d\lambda
\right].
\label{eq:drt}
\end{eqnarray}
\narrowtext
Spectral distributions of the intensities in the second and third terms are 
mainly located in the wavelength range of our observations (about 90\%). Only 
the first one, connected with $a^{3}\Sigma_{g}^{+},v' \gets 
X^{1}\Sigma_{g}^{+},v$=0 excitation needs the extrapolation described above. 
Therefore, the total dissociation rate in our conditions may be calculated 
from the measured continuum intensity. The results are shown in 
Fig.~\ref{fig:RMW}(a,b) for various pressures and ratios of the components in 
the mixture. They may be interpreted only qualitatively.

Figure~\ref{fig:RMW}(a) shows that the dissociation rate in the volume of 
observation monotonically decreases with increasing pressure.
The effect is analogous to that of the excited argon populations (see 
Fig.~\ref{fig:mw3a} and corresponding discussion). It is caused by the 
mechanisms already mentioned above, mainly by the decrease of microwave power 
coming to the volume of observation.

The observed changes in the dissociation rate caused by the variation of 
mixture components under constant pressure [Fig.~\ref{fig:RMW}(b)] are not 
trivial. In the absence of secondary effects one should expect linear 
dependence, whereas our measurements show a non-linear decrease of 
$(d[\mbox{H$_{2}$}]/dt)_{\text{diss}}$ towards smaller values of [H$_{2}$]. 
Most probably the effect is connected with very simple reasons, which can be 
of importance for industrial systems. When the content of hydrogen in the 
input gas mixture is high enough, the concentration of hydrogen in plasma is 
proportional to that in the input gas mixture. If the hydrogen content in the 
feed gas is reduced, then the hydrogen density in plasma is mainly influenced 
by resorption of hydrogen and water from the rather large metal 
surfaces.\footnote{The concentration of desorbed water vapor has been measured 
by IR absorption spectroscopy and was found to be some 10$^{14}$ molecules 
cm$^{-3}$, see \cite{RMKFD98}.} If it would be necessary to achieve small 
concentrations of hydrogen in plasma, then heating and pumping of the reactor 
must be done for a very long time.

On the other hand the data presented in Fig.~\ref{fig:RMW}(b) show that in our 
conditions the Ar*$\to$H$_{2}$ excitation transfer plays noticeable role even 
in the total rate of hydrogen dissociation. This effect should be taken into 
account in kinetics modeling of atomic hydrogen concentration in 
plasmachemical systems where Ar is used for transportation of reagents as a 
buffer. Our observation shows that it is not just a buffer.

\section{Conclusion}

In this paper for the first time the H$_{2}$ radiative dissociation continuum 
was used as a source of information about parameters of non-equilibrium 
plasma. Two new methods of spectroscopic diagnostics of hydrogen containing 
plasma have been developed: (1) determination of the vibrational temperature 
of the H$_{2}$ ground state from the relative intensity distribution and (2) 
derivation of the rate of electron impact dissociation from the absolute 
intensity of the continuum. The development of these methods has been based on 
a detailed analysis of the excitation-deactivation kinetics, rate constants of 
various collisional and radiative transitions as well as the way of data 
processing. The known method of vibrational temperature determination 
\cite{LP85} using H$_{2}$ emission line intensities of Fulcher-$\alpha$ bands 
was significantly improved and simplified. The potential of the new methods 
for plasma diagnostics was demonstrated at pure H$_{2}$ capillary-arc and 
H$_{2}$+Ar microwave discharges.

In pure hydrogen discharge plasma it was observed for the first time that for 
high enough current densities $j\ge3$~A~cm$^{-2}$ the shape of the continuum 
intensity distribution is influenced by the vibrational temperature of the 
ground $X^{1}\Sigma_{g}^{+}$ electronic state. Two independent spectroscopic 
methods gave almost the same values of $T_{\text{vib}}\approx 2000-5000$~K. 
The difference between vibrational and gas temperatures is in accordance with 
previous observations \cite{Capitelli86}.

In the H$_{2}$+Ar microwave plasma for the first time the shape of the H$_{2}$ 
dissociation continuum was found to depend on the mixture components, 
significantly influenced by Ar$^*\to$H$_{2}$ excitation transfer processes.
Absorption measurements of the population of the 3s$^2$3p$^5$4s levels of Ar 
together with certain computer simulation has been successfully used for 
verification of the proposed mechanism. Also for the first time the rates of 
radiative and total electron impact dissociation were obtained for various gas 
pressures and relative concentrations of the mixture components.

Our attempt to study the informational content of the intensity of the 
hydrogen dissociation continuum and to check its applicability for plasma 
diagnostics has been \mbox{\textit{a priory}} limited in two different 
respects.

1) We intended to have an experimental technique as simple as possible and, in 
particular, to stay on the basis of emission spectroscopy only. Then the 
methods would be especially useful in various industrial applications 
including rather easy \textit{in situ} control of plasma processing. The 
approach has in principle a backside because the knowledge of the actual 
electron energy distribution function and its spatial distribution may be of 
importance for a correct extraction of the information about plasma 
parameters. Langmuir probe measurements of electron energy distributions are 
not a problem nowadays, even in the cases of space-charge layers \cite{BLP87}, 
RF \cite{GPA92,FNWKK96} and microwave discharges (see \cite{KSTZ94} and 
Refs.~therein). We limited ourselves with intensity measurements in the more 
or less easy detectable near-UV region because the following second 
restriction of our attempt seems to be more significant.

2) Our analysis of the known characteristics of radiative and collisional 
processes involved into the mechanism of formation of observable  continuum 
emission led us to the conclusion that in spite of numerous efforts the 
situation is still unsatisfactory. Therefore, only simple models can be used. 
The corona-like model of the excitation, neglecting the rotational structure 
of upper vibronic levels, and the Franck-Condon approximation for electron 
impact excitation are actually available and justified right now.

Nevertheless, even in such complicated conditions we were happy to find that 
more or less accurate measurements of relative and absolute spectral 
distributions of the continuum intensity may be used for certain estimations 
of the vibrational temperature, for the analysis of the excitation mechanisms 
and for getting information about the rate of hydrogen dissociation by 
electron impact.

\begin{figure}
 \epsfxsize=8.6cm \epsfbox{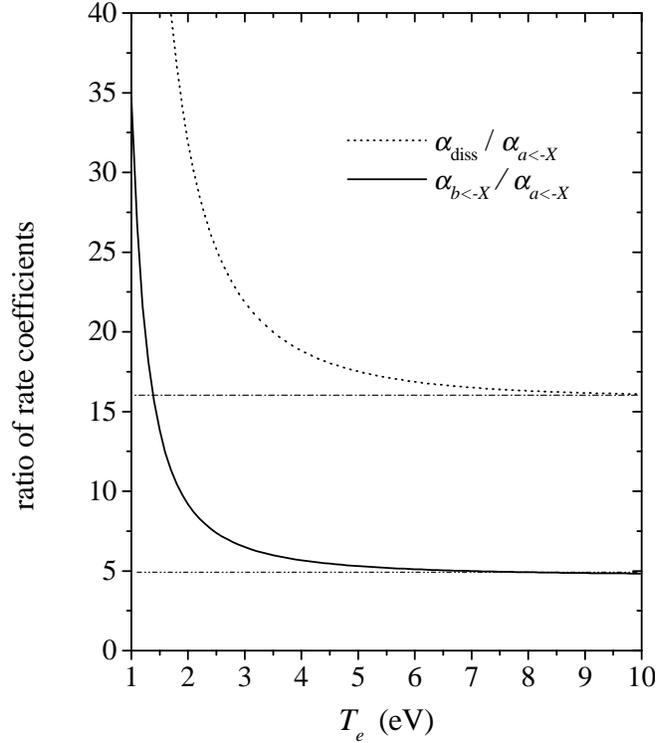}
 \caption{Ratios of the rate coefficients of electron impact excitation of 
$b^{3}\Sigma_{u}^{+}$ ($\alpha_{b\leftarrow X}$), $a^{3}\Sigma_{g}^{+}$ 
($\alpha_{a\leftarrow X}$) states and of the total electron impact 
dissociation ($\alpha_{\text{diss}}$) calculated as a function of the electron 
temperature.}
 \label{fig:rc}
\end{figure}

The last one looks more prospective because in some cases it can be already 
used as it is. If, for example, one is interested in the relative dependence 
of the electron impact dissociation rate on the position in plasma or on the 
discharge conditions, the information may be easily obtained just by a 
measurement of the continuum emission intensity even without any calibrations. 
The method of the determination of the absolute rate of the electron impact 
dissociation may be considerably improved, if the electron energy distribution 
function would be determined experimentally or calculated somehow. Then there 
would be no need to use our graph in Fig.~\ref{fig:rc} but to calculate the 
values in an explicit way by Eq.~(\ref{eq:alpha}).

This rather easy and independent method of determination of the absolute 
values of the rate of electron impact dissociation of hydrogen may be used in 
gas discharge physics and in plasma chemistry within two approaches.

1) It may be used for the determination of the hydrogen dissociation degree 
(which is often considered as a very important plasma parameter), if one is 
able to determine the effective mean lifetime of atoms in plasma with certain 
theoretical model of diffusion and association of atoms (see e.g.~\cite{LS87}).

2) If the value of the dissociation degree is determined by one of the 
existing methods (see \cite{L77,SvdGD96}) then the effective lifetime of atoms 
in plasma may be determined with the help of the rate of dissociation and the 
corresponding balance equation for atoms.\footnote{Laser-induced fluorescence 
spectroscopy using two-photon excitation is now being planned to determine the 
concentration of hydrogen atoms under the same conditions in plasma. In 
combination with the results derived in present paper by emission spectroscopy 
new channel of information about the lifetimes of hydrogen atoms in plasma can 
be opened.}

In both cases the method proposed in the present work is giving new 
opportunities.

Further improvements of the methods developed in present work need more 
detailed information about the elementary processes especially those about the 
cross sections of electronic-vibrational (better electronic-vibro-rotational) 
$a^{3}\Sigma_{g}^{+},v',N' \gets X^{1}\Sigma_{g}^{+},v,N$  electron impact 
excitation and the rate coefficients of the quenching of 
$a^{3}\Sigma_{g}^{+},v',N'$ levels in collisions with electrons, atoms and 
molecules. We certainly understand those are not simple problems, but we may 
hope that the needs of plasma diagnostics and our present work may be 
considered as some kind of the challenge for our colleagues working in the 
field of electronic and atomic collision physics.

\acknowledgements

The authors are indebted to Prof.~J.~Conrads, Prof.~S.~N.~Manida and 
Dr.~M.~Schmidt for general support of the project. This work was supported, in 
part, by the Deutsche Forschungsgemeinschaft, Sonderforschungsbereich 198, by 
the Russian Foundation for Support of Basic Research (grant No 95-03-09394a), 
and by the Program of Support of International Collaboration of Ministry of 
High Education of the Russian Federation. Dr.~A.~S.~Melnikov is thankful to 
the Deutsche Forschungsgemeinschaft, Sonderforschungsbereich 198, who provided 
a scholarship for his training in Germany and to the INP Greifswald for its 
hospitality. Prof.~B.~P.~Lavrov greatly appreciates INP Greifswald's financial 
support and hospitality. The authors are thankful to Prof.~N.~Sadeghi 
(Grenoble) for the discussion which stimulated the second and the third 
fittings presented in Table~\ref{tab:RMW}. One of us (BPL) is thankful to 
Prof.~V.~N.~Ostrovsky (St.~Petersburg State University) who informed him about 
the work \cite{LMBM91}. The authors thank Mr.~D.~G\"ott for skillful technical 
assistance.

% Figures

\clearpage

% Tables

\clearpage

\widetext
\begin{table}
\caption
{
Spectral distributions of transition probability for 
$a^{3}\Sigma_{g}^{+},v'$=0-6$\to b^{3}\Sigma_{u}^{+}$ spontaneous emission of 
the H$_{2}$ molecule and relative continuum intensity calculated for 
$T_{\text{vib}}\to0$ by Eq.~(\ref{eq:Il00}).
}
\begin{tabular}{cdddddddc}
$\lambda$&\multicolumn{7}{c}{$A_{b}^{av'}(\lambda)$ (10$^{12}$ cm$^{-1}$
s$^{-1}$) \ \ \ ~\protect\cite{LLP88}}&$J_{ab}(\lambda,355.5,0)$\\
\cline{2-8}
 [nm] &   $v'$=0 &   1 &   2 &   3 &    4 &   5 &    $v'$=6&present work\\
\tableline
160 &   0    &   0    &   0    &  3.220 & 49.674 &  3.795 & 22.977 & 3.717\\
170 &   0    &   0    &  1.714 & 34.647 &  9.474 & 23.008 & 14.436 & 5.339\\
180 &   0    &  0.395 & 16.152 & 24.766 & 11.127 &  0.144 &  2.297 & 6.034\\
190 &   0    &  3.701 & 26.715 &  0.231 & 10.235 & 10.524 &  5.725 & 6.454\\
200 &  0.283 &  9.938 & 17.191 &  6.123 &  0.047 &  2.873 &  5.463 & 6.147\\
210 &  0.949 & 14.879 &  3.565 &  9.571 &  4.340 &  0.697 &  0.077 & 5.590\\
220 &  2.252 & 14.894 &  0.016 &  2.799 &  6.257 &  4.182 &  1.898 & 4.829\\
230 &  3.851 & 11.283 &  2.370 &  0.377 &  3.070 &  4.066 &  3.662 & 4.300\\
240 &  5.315 &  6.723 &  4.702 &  0.386 &  0.386 &  1.681 &  2.498 & 3.666\\
250 &  6.540 &  3.087 &  5.315 &  1.871 &  0.164 &  0.135 &  0.704 & 3.218\\
260 &  7.025 &  0.948 &  4.222 &  3.010 &  1.154 &  0.165 &  0.025 & 2.788\\
270 &  7.090 &  0.085 &  2.645 &  3.056 &  1.971 &  0.851 &  0.262 & 2.407\\
280 &  6.792 &  0.049 &  1.330 &  2.440 &  2.176 &  1.416 &  0.797 & 2.089\\
290 &  6.249 &  0.389 &  0.463 &  1.598 &  1.918 &  1.609 &  1.161 & 1.822\\
300 &  5.582 &  0.819 &  0.079 &  0.893 &  1.414 &  1.456 &  1.256 & 1.616\\
310 &  4.867 &  1.224 &   0    &  0.367 &  0.869 &  1.126 &  1.118 & 1.451\\
320 &  4.165 &  1.509 &  0.108 &  0.101 &  0.471 &  0.755 &  0.866 & 1.325\\
330 &  3.530 &  1.671 &  0.295 &  0.006 &  0.197 &  0.451 &  0.597 & 1.222\\
340 &  2.946 &  1.748 &  0.474 &  0.020 &  0.045 &  0.226 &  0.372 & 1.130\\
350 &  2.453 &  1.717 &  0.618 &  0.100 &  0.005 &  0.080 &  0.193 & 1.042\\
360 &  2.021 &  1.647 &  0.695 &  0.202 &  0.011 &  0.018 &  0.083 & 0.955\\
370 &  1.667 &  1.530 &  0.818 &  0.296 &  0.058 &  0.002 &  0.024 & 0.885\\
380 &  1.373 &  1.399 &  0.846 &  0.379 &  0.117 &  0.016 &  0.004 & 0.809\\
390 &  1.125 &  1.265 &  0.844 &  0.445 &  0.179 &  0.047 &  0.007 & 0.736\\
400 &  0.921 &  1.127 &  0.818 &  0.486 &  0.228 &  0.088 &  0.026 & 0.665\\
410 &  0.735 &  0.985 &  0.781 &  0.499 &  0.271 &  0.127 &  0.051 & 0.592\\
420 &  0.621 &  0.865 &  0.729 &  0.504 &  0.303 &  0.161 &  0.080 & 0.534\\
430 &  0.514 &  0.752 &  0.674 &  0.497 &  0.320 &  0.190 &  0.104 & 0.477\\
440 &  0.423 &  0.654 &  0.615 &  0.477 &  0.327 &  0.207 &  0.127 & 0.424\\
450 &  0.343 &  0.567 &  0.559 &  0.455 &  0.327 &  0.221 &  0.144 & 0.376\\
460 &  0.283 &  0.488 &  0.503 &  0.429 &  0.322 &  0.226 &  0.157 & 0.333\\
470 &  0.241 &  0.420 &  0.452 &  0.397 &  0.310 &  0.229 &  0.164 & 0.295\\
480 &  0.203 &  0.363 &  0.402 &  0.366 &  0.297 &  0.229 &  0.168 & 0.261\\
490 &  0.169 &  0.312 &  0.359 &  0.337 &  0.281 &  0.221 &  0.168 & 0.231\\
\end{tabular}
\label{tab:Alv}
\end{table}
\narrowtext

\clearpage

\begin{table}
\caption
{
Radiative lifetimes of $d^{3}\Pi_{u}^{-},v,N$=1 levels and spontaneous 
emission transition probabilities of $d^{3}\Pi_{u}^{-},v,N$=1 $\to$ 
$a^{3}\Sigma_{g}^{+},v,N$=1 spectral lines of H$_{2}$.
}
\begin{tabular}{cr@{${}\pm{}$}lr@{${}\pm{}$}lr@{${}\pm{}$}l} 
%---------------------
  & \multicolumn{4}{c}{$\tau_{dv1}$ (ns)}
  & \multicolumn{2}{c}{$A^{dv1}_{av1}$ ($\mu$s$^{-1}$)}\\
\cline{2-5}
\cline{6-7}
$v$& 
\multicolumn{2}{c}{\cite{BLMPYY90,AKKKLOR96}\tablenotemark[1]} & 
\multicolumn{2}{c}{present work\tablenotemark[2]} &
\multicolumn{2}{c}{present work\tablenotemark[2]}\\
\tableline
0 & 40.7 & 1.4 & 38.7 & 2.0 & 24.4 & 1.4\\
1 & 38.4 & 1.3 & 39.7 & 2.1 & 20.6 & 1.1\\
2 & 39.5 & 1.9 & 40.9 & 2.2 & 16.9 & 1.0\\
3 & 39.5 & 0.9 & 42.2 & 2.2 & 13.6 & 0.8\\
4 & 19.0 & 1.0 & 43.6 & 2.4 & 10.6 & 0.6\\
5 & 15.2 & 1.2 & 45.5 & 2.5 & 8.1 & 0.5\\
6 & 16.0 & 3.0 &
\multicolumn{2}{c}{---}& 5.9 & 0.4\\
\end{tabular}
\tablenotetext[1]{Experiment.}
\tablenotetext[2]{Semiempirical calculation in the adiabatic approximation 
with dipole moment from \cite{LP88}.}
\label{tab:FtA}
\end{table}

%\widetext
\begin{table}
\squeezetable
\caption
{
Experimental and \textit{ab initio} data on the radiative lifetimes of 
$a^{3}\Sigma_{g}^{+},v'$ levels of H$_{2}$, p.~w. - present work.
}
\begin{tabular}{r@{${}\pm{}$}lr@{${}\pm{}$}lcddddc}
\multicolumn{9}{c}{$\tau_{av'}$ (ns)} &\\
\cline{1-9}
\multicolumn{2}{c}{$v'$=0}&\multicolumn{2}{c}{1}&2&3&4&5&$v'$=6& Ref.\\
\tableline
\multicolumn{9}{c}{calculated}&\\
\tableline
\multicolumn{2}{c}{11.9}&\multicolumn{2}{c}{11.0}
&  10.1   &  9.7  & --- & --- & ---&\cite{JC39}\\
\multicolumn{2}{c}{11.6}&\multicolumn{2}{c}{10.2}
&  9.17   &  8.40 & 7.81 & 7.30 & 6.90 &\cite{KGDP86}\\
\multicolumn{2}{c}{12.4}&\multicolumn{2}{c}{11.5}
&  10.3
& 9.29
& 8.64
& 8.07
& 7.63
& p.w.\\
\tableline
\multicolumn{9}{c}{experimental}                     &\\
\tableline
\multicolumn{9}{c}{\dotfill\,35$\pm$8\,\dotfill}
&\cite{FH65}\\
11.0 & 0.42  & 10.6 & 0.6  &---&---&---&---&---&\cite{IR71}\\
26 & 2&\multicolumn{2}{c}{---}&---&---&---&---&---&\cite{TF72}\\
11.9 & 1.2&10.8 & 1.1&
\multicolumn{2}{c}{\dotfill\,10$\pm$2\,\dotfill}
&---&---&---&
\cite{SC72}\\
\multicolumn{9}{c}{\dotfill\,10.45$\pm$0.25\,\dotfill}               
&\cite{KRI75}\\
9.94 & 0.39 & 9.1 & 1.0   &---&---&---&---&---&\cite{MK79}\\
\multicolumn{7}{c}{\dotfill\,9.62$\pm$0.20\,\dotfill}   &---&---&\cite{MK79}\\
\end{tabular}
\label{tab:tv}
\end{table}
\narrowtext

\begin{table}
\caption
{
Relative cross sections for electron impact excitation (in the maximum) of the 
$d^{3}\Pi_{u}^{-},v',N'$=1 levels from $X^{1}\Sigma_{g}^{+},v$=0,$N$=1 level 
and corresponding ratios of Franck-Condon factors.
}
\begin{tabular}{ccccclc}
$v'$&
\multicolumn{5}{c}{$\sigma^{dv'1}_{X01}/\sigma^{d21}_{X01}$}
&FCF\\\cline{2-6}
&\cite{MDH76}&\cite{BN76}&\hfill\cite{LOU81}\hfill&\hfill\cite{LP85}\hfill& 
present work& \cite{LMPT90}\\
\tableline
0 & 0.57 & 0.68 & 0.52$\pm$0.07 & 0.66$\pm$0.07 & 0.70$\pm$0.07 & 0.55\\
1 & 0.86 & 1.09 & 0.91$\pm$0.13 & 0.88$\pm$0.09 & 0.96$\pm$0.09 & 0.95\\
2 & 1.00 & 1.00 & 1.00$\pm$0.14 & 1.00$\pm$0.10 & 1.00$\pm$0.10 & 1.00\\
3 & 0.78 & 0.86 & 0.96$\pm$0.14 & 0.77$\pm$0.08 & 0.84$\pm$0.08 & 0.86\\
4 &  --- &  --- &  ---     & 0.35$\pm$0.04 & 0.87$\pm$0.09 & 0.66\\
5 &  --- &  --- &  ---     & 0.19$\pm$0.02 & 0.60$\pm$0.06 & 0.47\\
6 &  --- &  --- &  ---     & 0.11$\pm$0.01 & 0.32$\pm$0.03 & 0.33\\
\end{tabular}
\label{tab:2}
\end{table}

\begin{table}
\squeezetable
\caption
{
Franck-Condon factors for $a^{3}\Sigma_{g}^{+},v'\leftarrow 
X^{1}\Sigma_{g}^{+},v$ transitions of H$_{2}$ \protect\cite{LMPT90}.
}
\begin{tabular}{cccccccc}
&\multicolumn{7}{c}{$Q_{Xv0}^{av'0}$}\\
\cline{2-8}
$v'$    &$v=0$&     1&      2&      3&      4&      5&      $v=6$\\
\tableline
0&      0.20761&0.39958&0.28411&0.09341&0.01439&0.00089&0.00001\\
1&      0.25478&0.06483&0.08307&0.32592&0.21741&0.05018&0.00378\\
2&      0.20249&0.00503&0.16438&0.00281&0.20634&0.30686&0.10324\\
3&      0.13470&0.06063&0.05177&0.08235&0.07231&0.07873&0.34105\\
4&      0.08235&0.09580&0.00050&0.10148&0.00727&0.12529&0.01180\\
5&      0.04840&0.09603&0.01522&0.04193&0.06678&0.00933&0.11680\\
6&      0.02803&0.07908&0.04137&0.00439&0.07005&0.01583&0.04615\\
\end{tabular}
\label{tab:FCF1}
\end{table}

\begin{table}
\squeezetable
\caption
{
Franck-Condon factors for $d^{3}\Pi_{u}^{-},v'\leftarrow 
X^{1}\Sigma_{g}^{+},v$ transitions of H$_{2}$ \protect\cite{LMPT90}.
}
\begin{tabular}{cccccccc}
&\multicolumn{7}{c}{$Q_{Xv0}^{dv'0}$}\\
\cline{2-8}
$v'$    &$v=0$&     1&      2&      3&      4&      5&      $v=6$\\
\tableline
0&      0.09995&0.29272&0.33797&0.19710&0.06160&0.00994&0.00070\\
1&      0.17029&0.16255&0.00036&0.16128&0.29619&0.16752&0.03860\\
2&      0.18029&0.02851&0.08310&0.10636&0.00914&0.23640&0.26345\\
3&      0.15451&0.00075&0.10888&0.00041&0.12133&0.02528&0.11911\\
4&      0.11833&0.02432&0.05755&0.03636&0.05553&0.04245&0.09099\\
5&      0.08505&0.05077&0.01334&0.06851&0.00096&0.08332&0.00043\\
6&      0.05899&0.06393&0.00002&0.05697&0.01538&0.03661&0.04690\\
\end{tabular}
\label{tab:FCF2}
\end{table}

\begin{table}
\caption
{
Results of $\chi^2$-minimisation of the experimental data shown on 
Fig.~\ref{fig:probab7} with various number of adjusted parameters 
($v'_{\text{max}}$+1).
}
\begin{tabular}{cddc}
$v'_{\text{max}}$ & $\chi^2$ &     $\rho_{\text{max}}$     & 
$\left(\frac{\Delta N_{av'}}{N_{av'}}\right)_{\text{max}}$   \\
\tableline
0  &         57.1    &     10$^{-8}$  &     0.093        \\
1  &         9.89    &     0.62    &     0.108        \\
2  &         2.10    &     0.73    &     0.124        \\
3  &         0.426   &     0.84    &     0.127        \\
4  &         0.308   &     0.98    &     0.833        \\
5  &         0.149   &     1.00    &     6.670        \\
6  &         0.159   &     1.00    &     2.820        \\
\end{tabular}
\label{tab:RDVS1}
\end{table}

\begin{table}
\caption
{
Experimental and calculated with $v'_{\text{max}}=3$ relative population 
densities for $T_{\text{vib}}\to0$ of the $a^{3}\Sigma_{g}^{+},v'$ vibronic 
states and relative intensities $\xi_{v'}$ [Eq.~(\ref{eq:xi})] in the range 
$\lambda$=230-400~nm. Pure hydrogen capillary-arc discharge $i$=100~mA - see 
Fig.~\ref{fig:probab7}. Error(s) in brackets correspond to last digit(s).
}
\begin{tabular}{cllll}
$v'$ & \multicolumn{2}{c}{$N_{av'}/N_{a0}$} & \multicolumn{2}{c}{$\xi_{v'}$} \\
\cline{2-3}\cline{4-5}
&expt.&calc.&expt.&calc.\\
\tableline
0 &   1.00(3) &1.00& 0.50(2) & 0.50\\
1 &   1.20(4) &1.14& 0.27(1) & 0.25\\
2 &   0.79(7) &0.81& 0.14(1) & 0.14\\
3 &   0.86(11) &0.52& 0.09(1) & 0.06\\
4 &---&0.28&---&0.03\\
5 &---&0.15&---&0.01\\
6 &---&0.08&---&0.01\\
\end{tabular}
\label{tab:RDVS2}
\end{table}

\begin{table}
\squeezetable
\caption
{
Relative contributions $c_i$ of various channels of the excitation of the 
continuum emission [Eq.~\protect(\ref{eq:drt})] in the microwave plasma with 
constant pressure $p$ = 0.5 mbar for various H$_{2}$+Ar mixtures and the rates 
of radiative dissociation $-(d[\mbox{H$_{2}$}]/dt)_{ab}$ in 10$^{16}$ 
cm$^{-3}$ s$^{-1}$. Error in brackets corresponds to last digit.
}
\begin{tabular}{cr@{{}:{}}lccccccc}
Model&$[\mbox{Ar}]$&$[\mbox{H$_{2}$}]$ &$c_1$&$c_2$&$c_3$&$\chi^2$& 
&\multicolumn{2}{c}{$-(d[\mbox{H$_{2}$}]/dt)_{ab}$}\\
\cline{9-10}
&\multicolumn{7}{c}{}
&(1)\tablenote{Integrated over $\lambda=225-400$ nm.} 
&(2)\tablenote{Total radiative dissociation rate with extrapolation for entire 
range of $\lambda$.}\\
\tableline\multicolumn{9}{c}{}\\
&1  & 4&0.90(4)&0.10(6)&0.00(3)&0.93&&1.40&3.4\\
&1  & 1&0.79(4)&0.21(6)&0.00(3)&1.04&&1.05&2.4\\
$c_1,c_2,c_3 \neq 0$
&4  & 1&0.74(5)&0.23(7)&0.03(3)&1.27&&0.81&1.8\\
&9  & 1&0.72(5)&0.25(7)&0.03(4)&1.66&&0.78&1.7\\
&19 & 1&0.77(6)&0.18(8)&0.05(4)&1.90&&0.70&1.6\\
\tableline\multicolumn{10}{c}{}\\
&1  & 4&0.90(4)&0.10(4)&---&0.89&&1.37&3.3\\
$c_1,c_2\neq0$
&1  & 1&0.79(4)&0.21(4)&---&0.98&&1.05&2.4\\
%$c_1,c_2\neq0;\, c_3=0$
and
&4  & 1&0.73(5)&0.27(5)&---&1.26&&0.82&1.8\\
$c_3=0$
&9  & 1&0.71(5)&0.29(5)&---&1.64&&0.78&1.7\\
&19 & 1&0.75(6)&0.25(6)&---&1.96&&0.70&1.6\\
\tableline\multicolumn{10}{c}{}\\
&1  & 4&0.97(3)&---&0.03(2)&1.09&&1.36&3.4\\
$c_1,c_3\neq0$
&1  & 1&0.93(3)&---&0.07(3)&1.68&&1.06&2.6\\
%$c_1,c_3\neq0;\, c_2=0$
%$c_1,c_3\neq0;\, c_2=0$
and
&4  & 1&0.90(4)&---&0.10(3)&2.10&&0.82&2.0\\
$c_2=0$
&9  & 1&0.89(4)&---&0.11(3)&2.58&&0.79&1.9\\
&19 & 1&0.90(4)&---&0.10(3)&2.35&&0.71&1.7\\
\end{tabular}
\label{tab:RMW}
\end{table}
\end{document}